\documentclass[9pt,twocolumn,twoside]{pnas-new}
\templatetype{pnasresearcharticle} % Choose template 
\title{How directed is a directed network?}

\author[a,d,1]{R.S. MacKay}
\author[b,d]{S. Johnson}
\author[c]{B. Sansom}

\affil[a]{Mathematics Institute and Centre for Complexity Science, University of Warwick, Coventry, UK}
\affil[b]{School of Mathematics, University of Birmingham, Birmingham, UK}
\affil[c]{Mathematics and Economics, University of Warwick, Coventry, UK}
\affil[d]{The Alan Turing Institute, London, UK}

\leadauthor{MacKay} 

\significancestatement{In many domains of science, one is faced with a directed network and wishes to determine to what extent the edges line up in an overall direction.  The old concept of ``trophic level'' from ecology, and its more recent analogue ``upstreamness'' in economics, provided an answer but required basal or top nodes, respectively.  We introduce a new notion of trophic level and trophic coherence, which does not require basal or top nodes, is as easy to compute as the old notion, and is connected in the same way with network properties such as normality, cycles and spectral radius.  We expect this to be a valuable tool in domains from ecology and biochemistry to economics, social science and humanities.}

\authorcontributions{RSM obtained the grant for the project, came up with the improved notions of trophic level and incoherence and proved most of the results about them.  SJ did tests on a variety of networks, the comparisons of the new notion of trophic incoherence with other quantifiers of network structure, the ensemble theory for them, and some of the other proofs. BS built the Matlab toolbox, and did %the empirical analysis including 
many of the empirical tests reported here, including all those on supply networks and input-output networks. Each of us wrote parts of the text and contributed to its finalisation.}
\authordeclaration{The authors have no competing interests.}
\correspondingauthor{\textsuperscript{1}To whom correspondence should be addressed. E-mail: R.S.MacKay@warwick.ac.uk}


\begin{abstract}
The trophic levels of nodes in directed networks can reveal their functional properties. Moreover, the trophic coherence of a network, defined in terms of trophic levels, is related to  properties such as cycle structure,  stability and percolation. The standard definition of trophic levels, however, borrowed from ecology, suffers from drawbacks such as requiring source nodes, which limit its applicability. Here we propose a simple new definition of trophic level that can be computed on any directed network. We demonstrate how the method can identify node function in examples including ecosystems, supply chain networks, gene expression, and global language networks. We also explore how trophic levels and coherence relate to other topological properties, such as non-normality and cycle structure, and show that our method reveals the extent to which the edges in a directed network are aligned in a global direction.
\end{abstract}

\dates{This manuscript was compiled on \today}
\doi{\url{www.pnas.org/cgi/doi/10.1073/pnas.XXXXXXXXXX}}

\newcommand{\R}{\mathbb{R}}
\newcommand{\C}{\mathbb{C}}
\newcommand{\Z}{\mathbb{Z}}
\newcommand{\tr}{\mbox{tr}}
\newcommand{\diag}{\mbox{diag}}

\begin{document}

\maketitle

\thispagestyle{firststyle}
\ifthenelse{\boolean{shortarticle}}{\ifthenelse{\boolean{singlecolumn}}{\abscontentformatted}{\abscontent}}{}

% If your first paragraph (i.e. with the \dropcap) contains a list environment (quote, quotation, theorem, definition, enumerate, itemize...), the line after the list may have some extra indentation. If this is the case, add \parshape=0 to the end of the list environment.
\dropcap{M}any complex systems have an underlying network, whose nodes represent units of the system and whose edges indicate connections between the units \cite{N}.  In some contexts the connections are symmetric, but in many they are directed, for example indicating flows from one unit to another or which units affect which other units \cite{BG}.   
%Our interest here is in directed networks.

In a directed network the ecological concept of ``trophic level'' \cite{Levine} allows one to assign a height to each node in such a way that on average the height goes up by one along each edge.  
The trophic levels can help to associate function to nodes, for example, plant, herbivore, carnivore in a food web.
The concept was reinvented in economics \cite{ACFH}, where it is called ``upstreamness''.

The standard deviation of the distribution of height differences along edges gives a measure of the extent to which the directed edges line up, called the trophic incoherence \cite{JDDM}.
The trophic incoherence is an indicator of network structure that has been related to stability, percolation, cycles, normality and various other system properties \cite{JJ,J,KJ1, D+, KJ2}.

The standard definitions of trophic level and incoherence are limited in various ways, however.  In particular, they require the network to have basal nodes (source nodes), they give too much emphasis to basal nodes if there is more than one, they are not symmetric with respect to reversing all the edges, they do not give a stable way to determine levels and incoherence for a piece of a network, and 
%although they give a natural notion of maximal coherence 
they do not give a natural notion of maximal incoherence.

In this paper we present improved definitions of trophic level and incoherence that overcome these limitations.  We illustrate their application in a variety of domains.  We show that the new levels continue to be a useful indicator of function in the network and that the new incoherence measure continues to be related to stability, cycles and normality.  We compare the new notion with the old for cases that have basal nodes.  And we show the robustness of our new trophic levels to truncation of a network.

\newpage
\section{The new notions of trophic level and incoherence}
We consider directed networks (also known as directed graphs or digraphs) with set $N$ of nodes (also known as vertices) and set $E$ of directed edges (also known as links).
We suppose that there is at most one edge from a node $m$ to a node $n$, and denote the edge by $mn$.  There can also be an edge from $n$ to $m$.  Each edge carries a weight
$w_{mn}>0$.  This can represent the strength of the edge.  We write $w_{mn}=0$ if there is no edge from $m$ to $n$ and we assemble the $w_{mn}$ into a matrix $W$.  The edge weights could be set to $1$, as is common in the literature, and the array $W$ is then called the adjacency matrix $A$ of the network, but the ability to represent the strength of the edge is a useful extension.
If there were multiple edges from $m$ to $n$ then we would amalgamate them into a single edge by adding the weights.  Self-edges $mm$ (also called loops) are permitted.  

For each node $n$ we define its in-weight and out-weight by
\begin{equation}
w_n^{in} = \sum_{m\in N} w_{mn}, \quad w_n^{out} = \sum_{m\in N} w_{nm}.
\end{equation}
We define the {\em weight} of the node $n$ by
\begin{equation}
u_n = w_n^{in}+w_n^{out},
\label{eq:node_weight}
\end{equation}
and the {\em imbalance} for node $n$ by
\begin{equation}
v_n = w_n^{in}-w_n^{out}.
\end{equation}
The (weighted) {\em graph-Laplacian operator} $\Lambda$ on functions $h:N \to \R$ is defined by
\begin{equation}
(\Lambda h)_m = u_m h_m - \sum_{n\in N} (w_{mn} +w_{nm})h_n,
\end{equation}
or in matrix form (where $^T$ denotes transpose),
\begin{equation}
    \Lambda = \diag (u) - W-W^T.
\end{equation}

Then our improved notion of {\em trophic level} is the solution $h$ of the linear system of equations
\begin{equation}
\Lambda h = v .
\label{eq:Lhv}
\end{equation}
%(For an electrical interpretation of [\ref{eq:Lhv}], see the Supporting Information (SI).)

The equations [\ref{eq:Lhv}] always have a solution (see the Supporting Information (SI)) but it is non-unique, because one can add an arbitrary constant in each connected component of the network.  A {\em connected component} of a network is a maximal subset $S \subset N$ such that it is possible to get from any $m \in S$ to any $n \in S$ by a path of edges ignoring their directions.  
Thus to solve $\Lambda h = v$ one can replace the equation for one node $m_S$ in each connected component $S$ by an equation $h_{m_S}=c_S$ for arbitrary constants $c_S$, for example $0$.  Then there is a unique solution for $h$, which can be found by any linear algebra package.  Afterwards one can add an arbitrary constant to the levels in each component $S$ if desired, for example to make the lowest one be $0$ or to make the average level (with respect to the weights $u_n$, for example) in $S$ be $0$.

Our improved notion of {\em trophic incoherence} is
\begin{equation}
F_{0} = \frac{\sum_{mn} w_{mn}(h_n-h_m-1)^2}{\sum_{mn} w_{mn}}.
\end{equation}
This has the nice features that $F_0 = 0$ if and only if all the level differences $z_{mn} = h_n - h_m$ are $1$, $F_0 =1$ if and only if all the level differences are $0$, and otherwise $F_0$ is strictly between $0$ and $1$ (see SI for a proof).  We say a network is {\em maximally coherent} if it has $F_{0}=0$, {\em maximally incoherent} if it has $F_{0}=1$.  We define the {\em trophic coherence} to be $1-F_0$.  In the SI we prove the trophic coherence can be expressed alternatively as the weighted mean difference $\bar{z}$ in trophic levels between nodes along the edges of the network.

The motivation for our new definitions is to seek levels $h_n, n \in N,$ that minimise the {\em trophic confusion}
\begin{equation}
F(h) = \frac{\sum_{mn} w_{mn}(h_n-h_m-1)^2}{\sum_{mn} w_{mn}},
\end{equation}
where the target level difference for each edge $mn$ is set to $1$.  A vector $h$ of levels minimises $F$ if and only if $\Lambda h = v$ (see SI).  The resulting minimum value of $F$ is $F_0$.

\section{Illustrations}
\begin{figure}[htb] %  figure placement: here, top, bottom, or page
   \centering
   \includegraphics[width=8.0cm]{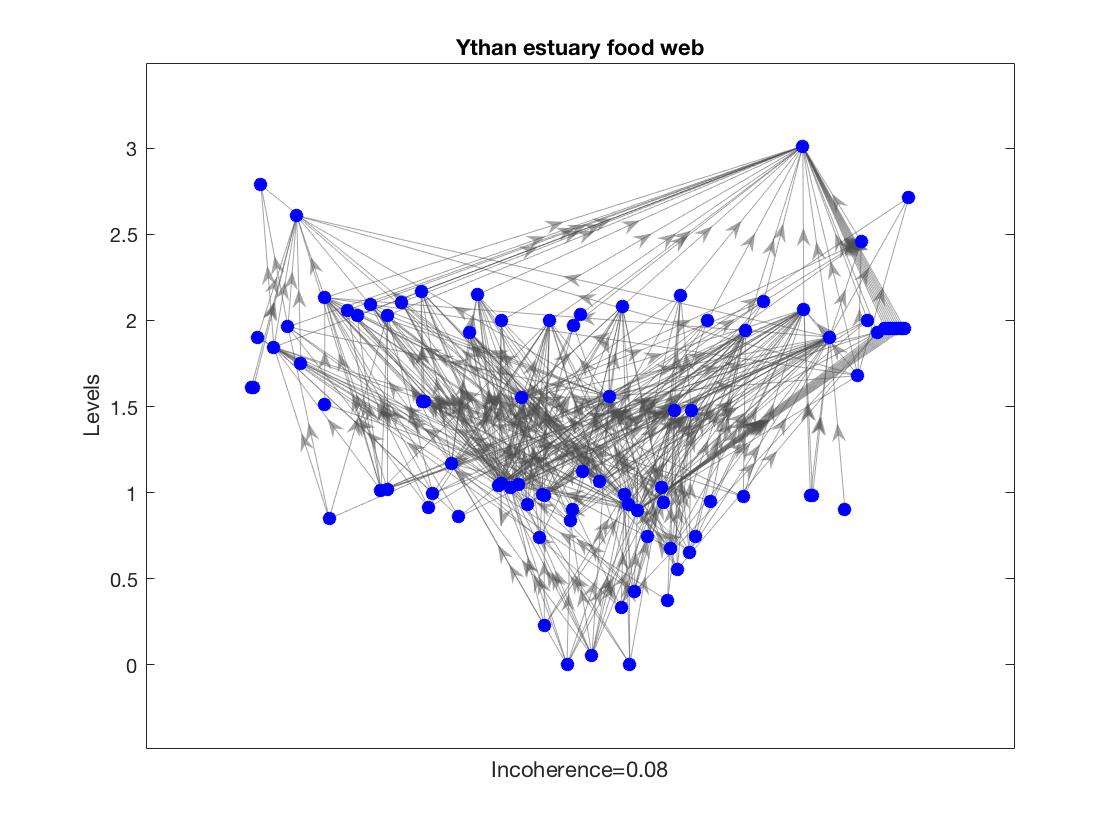} 
   \caption{Ythan estuary food web with height corresponding to our new trophic levels which reveal a strongly layered structure. Edges represent prey$\rightarrow$predator relations and edge weights are all taken to be 1 as the data does not specify the relative importances of relationships.}%Stony Stream food web laid out with height corresponding to our new trophic levels.}
   \label{fig:Ythan}
\end{figure}

To illustrate the new notions of trophic level and incoherence, 
we begin with the classic context of food webs.  Here the nodes are species and there is an edge from a species to each species that eats it.  %The edge weights are all taken to be 1 if the data does not specify the relative importances of the trophic relationships.  
Figure~\ref{fig:Ythan} shows the Ythan estuary food web \cite{C+} with height in the layout corresponding to our new notion of trophic level. 
%One sees that there is a large set of species whose trophic level is about 0 and another large set of species whose height is about 1, and then a few at a range of higher levels (note that some of them are cannibals) but also one at a level of about 0.5. 
The network is fairly strongly layered;  this is borne out by a small value of  trophic incoherence $F_{0}=0.08$. 
% Stoney:($F_{0}=0.019$).

\begin{figure}[h!bt] %  figure placement: here, top, bottom, or page
   \centering
   \includegraphics[width=8.5cm]{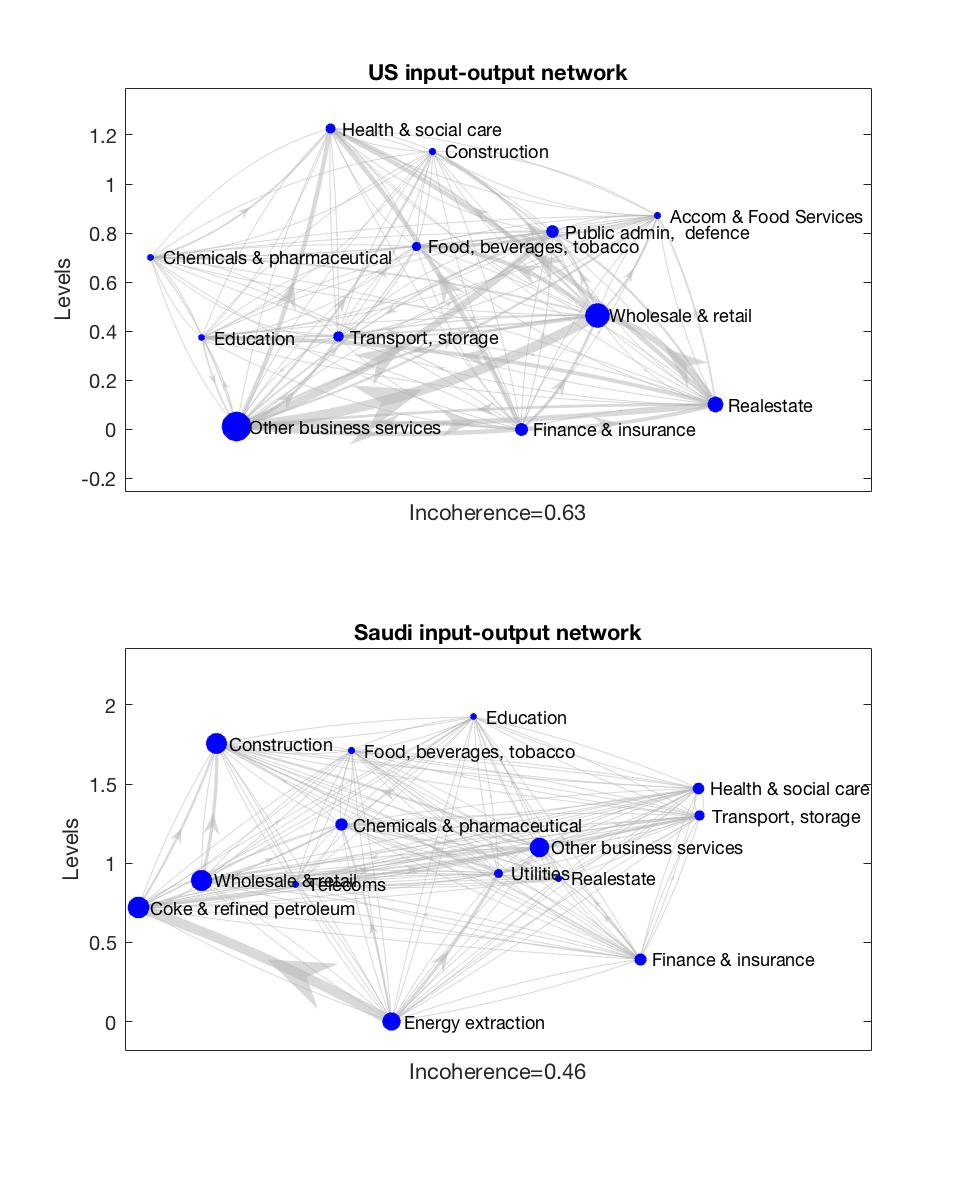} 
   \caption{Network of inter-industrial flows of goods and services in the US (top) and Saudi (bottom) economies in 2015. Nodes represent a subset of economic sectors (accounting for largest share of inter-industry flows as captured by {weight} [\ref{eq:node_weight}]) and weighted edges represent the dollar value of supply$\rightarrow$purchase transactions between them. Edge widths reflect the value of flows, and node size reflects node {weight} [\ref{eq:node_weight}].}
   \label{fig:IO}
\end{figure}
We continue with an example from economics where 
%e.g.~ 
the `upstreamness'/`downstreamness' of \textit{firms}, \textit{sectors} and \textit{economies} in production-chains is of wide relevance and interest \cite{AC,B, MSCF}.
Figure~\ref{fig:IO} shows the inter-industrial flows of goods and services in the US and Saudi economies in 2015 (data taken from OECD input-output (IO) tables). Here the nodes represent economic sectors and weighted edges represent the dollar value of supply$\rightarrow$purchase transactions between them %(data taken from OECD input-output (IO) tables). 
(the full IO table had 35 sectors, but nodes with lower {weight} [\ref{eq:node_weight}] were removed to allow presentation of a labeled network). This is an interesting application because there are no basal nodes (indeed the networks are fully connected, as is usual for IO relations, with every sector both supplying and buying from every other sector), so the old notions of trophic level and incoherence cannot be applied.

Unlike the Ythan food web, these IO networks are rather incoherent and this is borne out by much higher values of trophic incoherence ($F_{0}=0.63$ and $F_{0}=0.46$ respectively). Nevertheless the new levels reveal the overall direction of flow in intermediate production:~some sectors are key suppliers of intermediate inputs (for the US, financial, real estate and other business service sectors; for Saudi Arabia, energy extraction and finance) while other sectors are key users of inputs from other sectors (e.g.~healthcare and construction).

\begin{figure}[htb] %  figure placement: here, top, bottom, or page
   \centering
   \includegraphics[width=8.5cm]{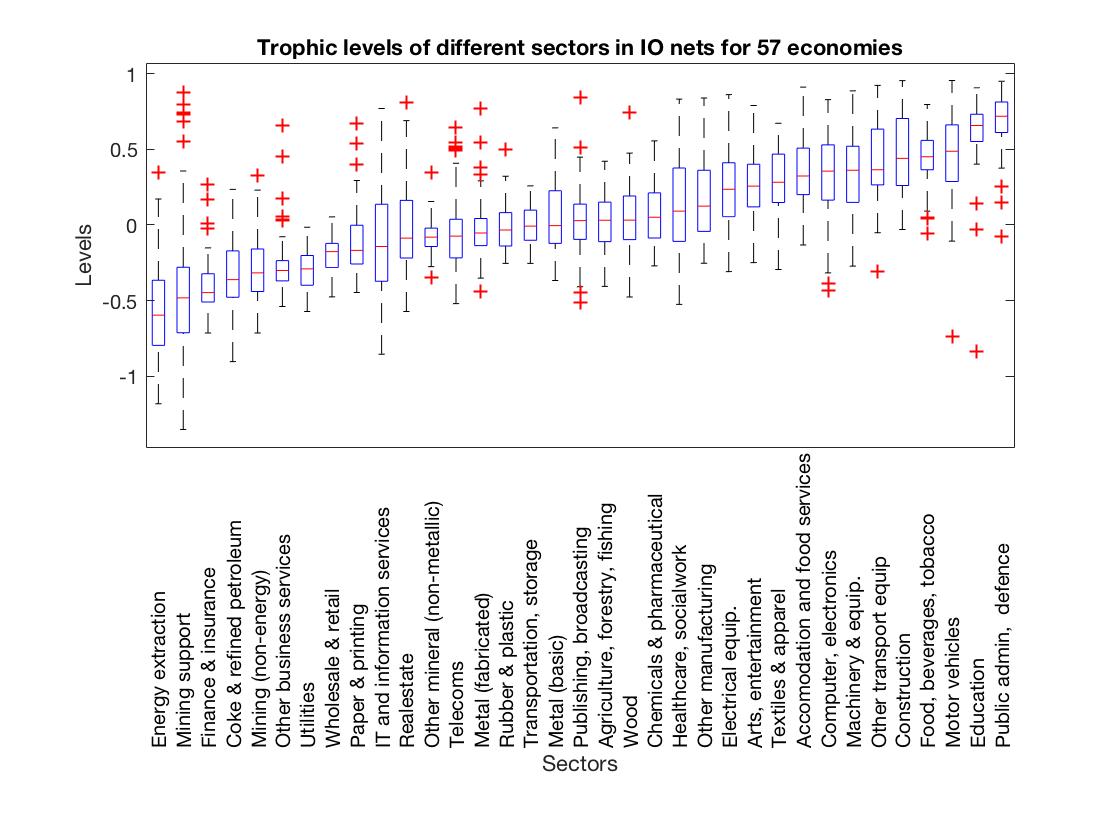}%[width=5in] 
   \caption{These boxplots present the distribution of the new trophic levels for each of 35 different economic sectors (ISIC Rev.~4) as obtained from the national 2015 input-output (IO) networks of 57 different economies (including OECD, and G20 economies). Sectors are sorted by their median trophic level (across all 57 IO networks). Red crosses indicate outliers.}
   %Although these are fully connected networks and rather incoherent (with $F_{0}$ ranging from 0.47 to 0.88), the trophic levels based on the weighted networks (according to the flows of goods and services between sectors) reveal the hierarchical architecture of value chains in these economies. IO data from OECD IO Tables.}
   \label{fig:sectors}
\end{figure}
%Input-output (IO) networks capture sale and purchase relationships between producers and consumers within an economy. 

%Figure~\ref{fig:poems} shows the word-graphs for two poems, %where the nodes represent words and an edge indicates that %one word is followed by another somewhere within a sentence.
%PERHAPS THESE ARE TOO SHORT TO REVEAL MUCH?
%\begin{figure}[htb] %  figure placement: here, top, bottom, or page
%   \centering
%   \includegraphics[width=8.5cm]{Fig2clip.png}%[width=5.5in] 
%   \caption{Word-graphs for two sonnets, with height indicating trophic level. 
%   Left: Shakespeare's Sonnet 18 ({\it Shall I compare thee to a summer's day?}).
%Right: Michelangelo's Rima 3. ({\it Grato e felice, a' tuo feroci mali}).
% }
 %  \label{fig:poems}
%\end{figure}

%It is interesting that the Italian poem is much more trophically coherent than the other.  Some correspondence can be seen between syntactic role and height?
%OTHER KEY COMMENTS BUT LESS DETAILED THAN IN THE PREVIOUS DRAFT

Figure~\ref{fig:sectors} provides a more systematic and detailed analysis, presenting box-plots of the {level} of different sectors (using full 35 sector IO tables) for 57 countries (2015 data). Levels for each economy have been normalised to make the mean level (weighted by $u_n$) 0. While the size of different sectors varies across economies, there is considerable consistency of sector levels, which reveal the hierarchical architecture of value chains in the production process:~we see an overall direction of flow from energy extraction and finance sectors; through other primary materials; then manufacturing industries; followed by sectors that supply final demand more directly, such as food makers, entertainment,  and services; ending  with education, public administration and defence sectors %(sectors 
(that are overwhelmingly users more than suppliers of intermediate inputs).

%\begin{figure}[htb] %  figure placement: here, top, bottom, or page
%   \centering
%   \includegraphics[width=8.5cm]{Fig4_sector_levels.jpg}%[width=5in] 
%   \caption{These boxplots present the distribution of the new trophic levels for each of 35 different economic sectors (ISIC Rev.~4) as obtained from the national input-output (IO) networks of 57 different economies (including OECD, and G20 economies). 
   %IO networks capture sale and purchase relationships between producers and consumers within an economy. 
%   Sectors are sorted by their median trophic level (across all 57 IO networks). Although these are fully connected networks and %the networks are 
%   rather incoherent (with $F_{0}$ ranging from 0.47 to 0.88), the trophic levels based on the weighted networks (according to the flows of goods and services between sectors) reveal the hierarchical architecture of value chains in these economies. IO data from OECD IO Tables.}
%   \label{fig:sectors}
%\end{figure}
%Input-output (IO) networks capture sale and purchase relationships between producers and consumers within an economy. 
%Although these are fully connected networks (every sector both supplies and buys from every other sector),
%with $F_0$ ranging from 0.47 to 0.88), 
%thus standard levels cannot be obtained, the new levels 
%based on the weighted IO network (edges are weighted according to the flows of goods and services between sectors) 
There may be links to explore between sector levels and their role in economic performance -  it is interesting for example to note that construction appears as a key user of inputs from other industries (implying strong backward-linkages) given the stylized business-cycle fact that house building leads the wider cycle \cite{Leamer}. Meanwhile variation in the level of some sectors across different economies may also reveal interesting differences in production structure (e.g.~finance occupies the same minimum position as energy extraction in China, but comes higher in the value chain for many other economies).

\begin{figure}[htb] %  figure placement: here, top, bottom, or page
   \centering
   \includegraphics[width=8.5cm]{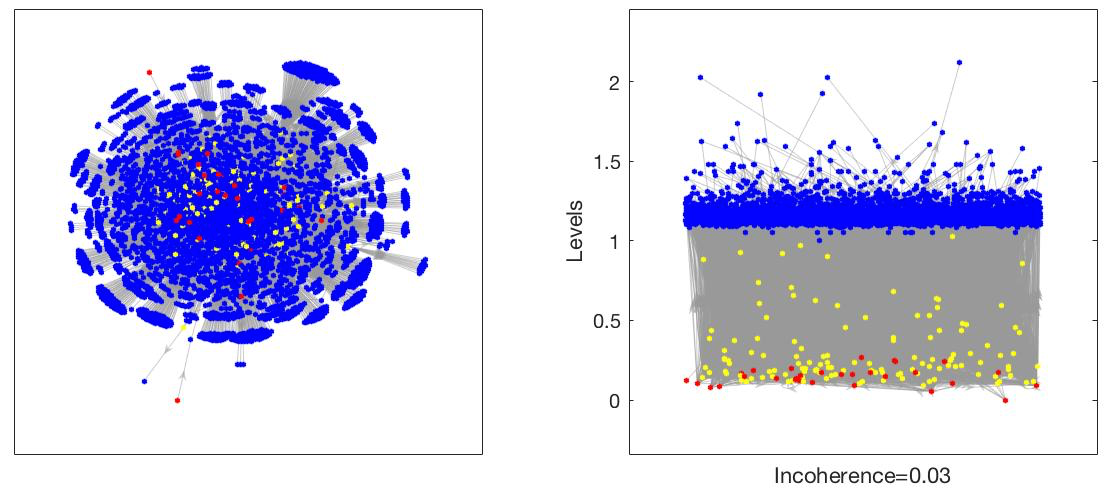}%[width=5.5in] 
   \caption{These two charts plot the same yeast transcription regulatory network (linking transcription factors and target genes) first with a standard force directed layout (left); then with node heights corresponding to new trophic levels $h$ to reveal the network's flow-based hierarchy. Red nodes represent transcription factors, blue nodes denote regulated genes, and those with both functions are coloured in yellow.
 }
   \label{fig:regulatory}
\end{figure}

In biology, regulatory networks are sets of macromolecules %mostly proteins and RNAs, 
that interact to control the level of expression of various genes in a given genome [\href{https://www.nature.com/subjects/regulatory-networks}{Nature subjects: Regulatory networks}]. %The main players in regulatory networks are DNA-binding proteins, also called transcription factors as they modulate the first step in gene expression. 
Studies on regulatory networks have identified the existence of hierarchical structures and linked node levels to node properties, function \cite{G,Jothi,Yu} and %evidence that hierarchy more than connectivity better reflects 
the importance of regulators \cite{BKG}. Assigning hierarchical levels in cyclic networks, however, has presented a methodological challenge for this literature which our new levels overcome. Figure~\ref{fig:regulatory} shows an example transcription regulatory network (the yeast \textit{Saccharomyces cerevisiae} \cite{Jothi}) plotted first with a force directed method (left), then according to the new levels (right). The new levels reveal a striking hierarchical structure. There are source (red), intermediate (yellow) and target (blue) nodes, but intermediate nodes do not form a distinct layer and the relevance of variation in their levels might be explored.

\begin{figure}[htb] %  figure placement: here, top, bottom, or page
   \centering
   \includegraphics[width=8.5cm]{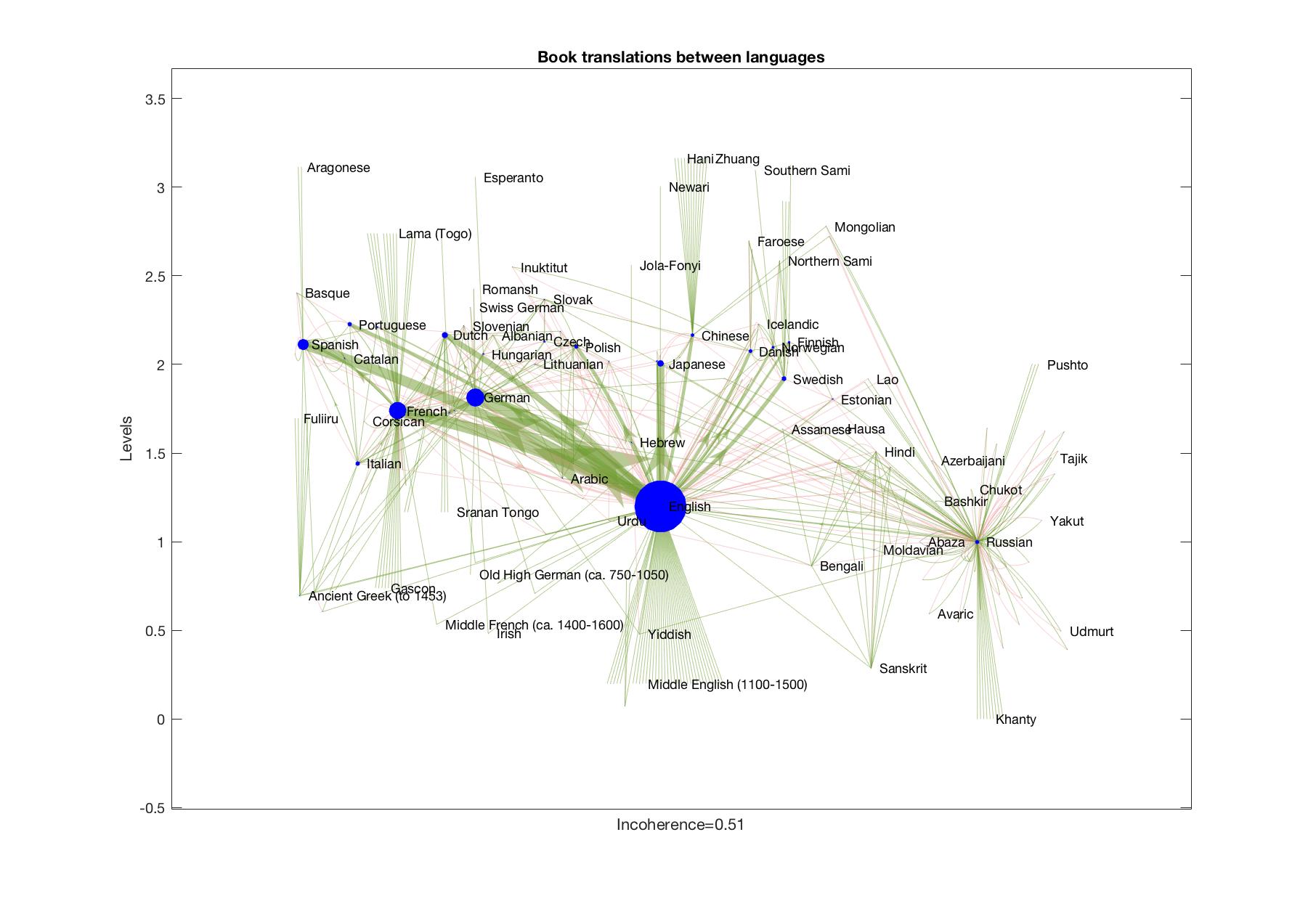}%[width=5.5in] 
   \caption{Global book translation network \cite{RG}. Edges and edge weights represent the number of books translated from source into target language. Upward arrows are plotted green and downward arrows red. Node size is proportional to {weight} [\ref{eq:node_weight}]).
 }
   \label{fig:books}
\end{figure}

Flow-based hierarchies may also be important in social network settings (hierarchy and stratification are important concepts in sociology) and have been studied in e.g.~online social networks \cite{Gu,Lu}.

Figure~\ref{fig:books} shows the trophic analysis of a network of book translations \cite{RG} based on a collection of more than 2.2 million book translations compiled by UNESCO’s \textit{Index Translationum project} \cite{unesco}. %(an international index of printed book translation \cite{unesco}). 
Edge weights correspond to the number of books translated between source and target languages. % (and represent at least six books) [max edge 183,329, min edge 6].
Our new levels reveal interesting information on the position of different languages in this global network: at the bottom appear languages that are only source languages - unsurprisingly these include many `dead languages' (Ancient Greek, Middle French and English, Sanskrit etc.). At the top appear languages nobody translates (these include minority and other languages that are small by number of speakers such as Faroese, Sami, and Mongolian). In the middle we find languages that are both target and source languages. The central role of English is striking: whilst translated into and out of, English is more important as a \textit{source} language (lower in the hierarchy than any other major languages) and there are large flows from English into French, German, Spanish and Japanese. In this data-set only English is translated into Chinese which is in turn only a source language for minority languages in China (such as Hani and Zhuang). Russian is rather isolated in the global language network but forms an interesting community of bi-directional links with languages in its region.

Overall the network is surprisingly coherent ($F_{0}=0.51$). While it is unlikely individual books flow along paths in this network (given books are presumably translated from original source language) its structure may be important in the flow of knowledge and ideas \cite{RG}, and trophic analysis helps shed light on the strongly hierarchical structure in the global language network.

\section{Comparison with old notions}
The established concept of trophic level \cite{Levine} requires the network to have at least one basal node, that is a node with no incoming edges.  Then the height $x_n$ (to use Levine's symbol) was set to a common value of $0$ for all basal nodes $n$, though nowadays it is more common to set it to $1$.  The heights of the other nodes in connected components with basal nodes were determined by solving
\begin{equation}
x_n = 1 + \frac{\sum_{m} x_m w_{mn}}{w_n^{in}}
%{\sum_m w_{mn}},
\label{eq:Levine}
\end{equation}
for all non-basal $n$, where each sum is over the nodes $m$ having edges to $n$. Levine normalised the weights $w_{mn}$ coming into each node $n$ so that $w_n^{in}=1$,
%$\sum_m w_{mn} = 1$,
which makes no change to [\ref{eq:Levine}].
%Actually, Levine took all edge weights $1$, but the above formula is a natural generalisation to allow arbitrary positive weights.   
In matrix form, the equation for the heights (with the convention $x_n=1$ for basal nodes) can be written as
\begin{equation}
Lx = \tilde{v}
\end{equation}
where 
\begin{equation}
\tilde{v}_n = w_n^{in} \mbox{ if non-zero, else } 1,
\end{equation}
and
\begin{equation}
(Lx)_n = \tilde{v}_n x_n - \sum_{m} x_m w_{mn} .
\end{equation}
The same concept was introduced in economics by \cite{ACFH}, but fixing top nodes (those with no outgoing edges) to a common height.  It is equivalent to Levine's after reversing all the edges.

Then \cite{JDDM} defined the trophic incoherence of the network to be the standard deviation of the height differences $z_{mn} = x_n-x_m$ over edges.  They took edge weights all $1$, but a natural generalisation is to weight the height differences by the edge weights.  The edge-weighted mean difference of Levine's heights is precisely $1$ \cite{Levine}, so Johnson et al's definition of trophic incoherence $q$ becomes
\begin{equation}
q = \sqrt{\frac{\sum_{mn} w_{mn} (x_n-x_m-1)^2}{\sum_{mn} w_{mn}}}.
\end{equation}
Indeed, Levine defined ``trophic specialisation'' of a node $m$ as
\begin{equation}
    \sigma^2_m = \frac{\sum_n w_{mn} (x_n-x_m-1)^2}{\sum_n w_{mn}}.
\end{equation}
So $q^2$ is the average of $\sigma^2_m$ weighted by
$w_m^{out}$.
%$\sum_n w_{mn}$.

Our equation for trophic heights can be seen as a symmetrised version of Levine's, without the fix for basal nodes.  Thus our definition doesn't need any basal nodes and does not force them all to the same level if there is more than one basal node.  

Our definition of trophic incoherence is the same as $q^2$ but using our new heights instead of Levine's.  It represents, in roughly the same way, the failure of the height differences to all be $1$.  A distinction to bear in mind, however, is that for our new levels, the edge-weighted mean height difference 
\begin{equation}
\bar{z} = \frac{\sum_{mn} w_{mn}(h_n-h_m)}{\sum_{mn} w_{mn}}
\label{eq:zbar}
\end{equation}
is not necessarily $1$.
In fact, we prove in the SI that $\bar{z} = 1 - F_0$.
So $F_{0}$ is not in general the (edge-weighted) variance of the height differences.  To obtain the variance $\sigma^2$ of the height differences one has to subtract $(\bar{z}-1)^2$ from $F_{0}$.  Thus there is a case for considering alternative measures of incoherence to $F_{0}$, such as the ratio $\eta = \sigma/\bar{z}$, which evaluates to 
\begin{equation}
\eta = \sqrt{\frac{F_0}{1-F_0}}
\label{eq_eta}
\end{equation}
and is the appropriate replacement for $q$. In the other direction, the analogue of $F_0$ is ${q^2}/{(1+q^2)}$.

\begin{figure}[htb] %  figure placement: here, top, bottom, or page
   \centering
   \includegraphics[width=8.5cm]{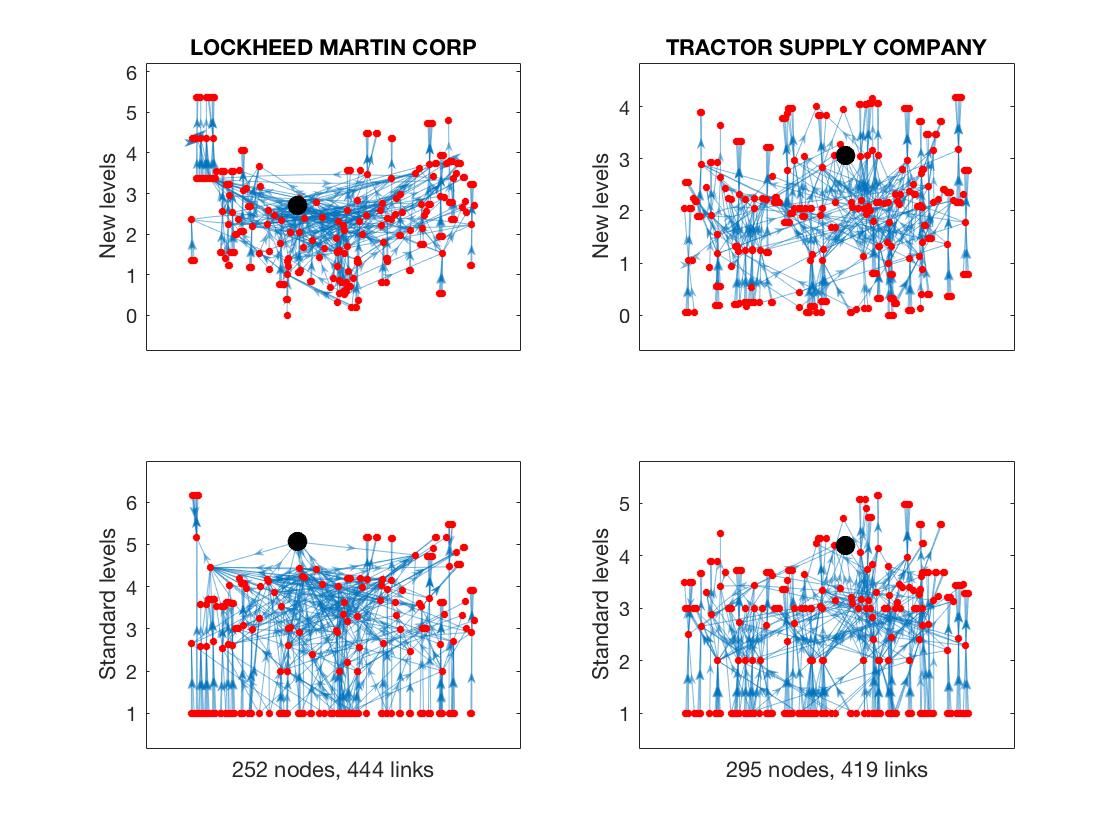} 
   \caption{Supply networks around two firms (indicated by the larger black dot), plotted with the new (top) and old (bottom) notions of trophic level. The horizontal positions are determined to spread out the nodes while attempting to make most of the edges near vertical, but the same horizontal positions are used in both the upper and lower pictures. Supply chain data compiled from Bloomberg L.P. }
   \label{fig:supply}
\end{figure}

Figure~\ref{fig:supply} shows some comparisons of trophic levels for two networks with basal nodes, determined by the two methods.  They are both supply networks, extracted from Bloom\-berg by taking all suppliers and buyers within 3 hops of a given firm (a hop being an edge in either direction).  The nodes represent firms and a directed edge represents that the first firm supplies goods or services to the second.  
We see that the requirement of the standard approach to put all basal nodes at a common level makes an artificial distortion of the levels in the lefthand case, though less so on the right.  %Indeed, when we extended the network to a larger neighbourhood, the levels moved significantly for the old notion but less so for the new one [CAN YOU CONFIRM THAT, BAZIL? BS: DON'T HAVE THIS DATA!]

\begin{figure}[htb] %  figure placement: here, top, bottom, or page
   \centering
   \includegraphics[width=8.5cm]{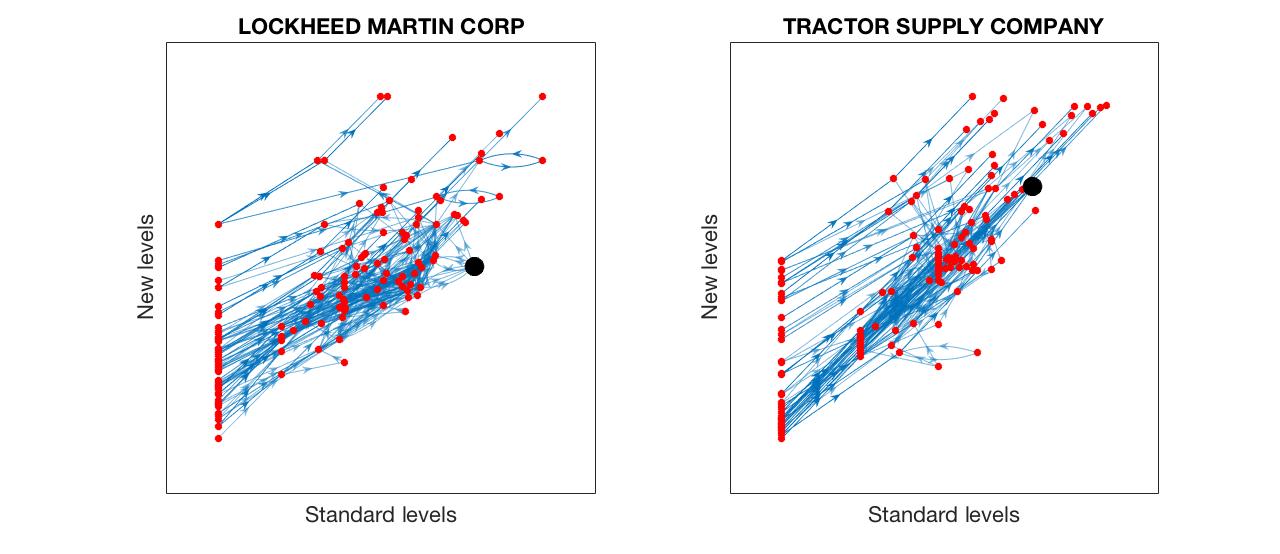} 
   \caption{New levels against old levels for the same two supply networks as in Figure~\ref{fig:supply}.}
   \label{fig:comparison}
\end{figure}

As an alternative comparison, in Figure~\ref{fig:comparison} we plot (for the same two supply networks as Figure~\ref{fig:supply}) the old levels against the new levels.

If one reverses all the edges then with our new definition one obtains the reflection of the trophic levels, up to an overall shift depending on the convention used to fix the zero of the levels.  The trophic incoherence is unchanged.
For example, for a supply network, instead of the flows of goods and services one could instead consider the flows of payment, which are more or less the reverses of the flows of goods and services.  

\begin{figure}[htbp] %  figure placement: here, top, bottom, or page
   \centering
   \includegraphics[width=8.0cm]{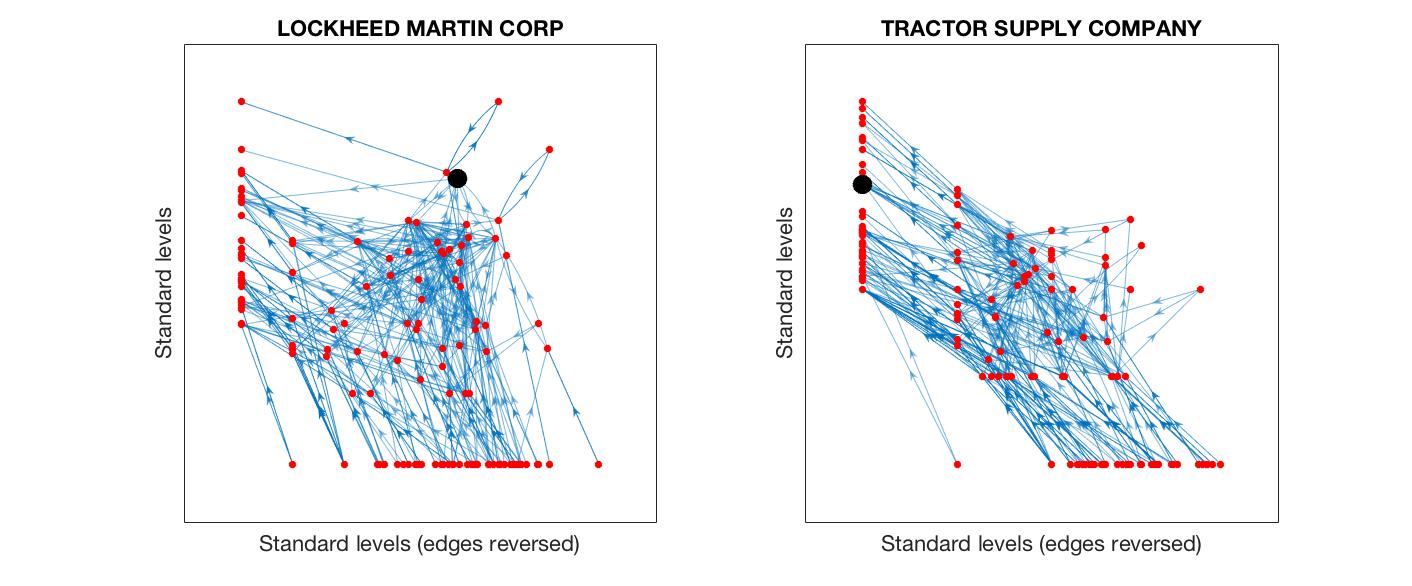} 
   \caption{The same two supply networks as in Figure~\ref{fig:supply} and ~\ref{fig:comparison} with nodes organised according to trophic levels obtained using the old notion for (i) the original networks (vertical axis) and (ii) the same networks but with interpretation of edges reversed (horizontal axis).}
   \label{fig:new}
\end{figure}

%Similarly, we demonstrate the failure of 
In contrast, the old notion of trophic level 
%to respect symmetry 
is usually not symmetric with respect to change of direction of all the edges.  Figure~\ref{fig:new} shows the trophic levels of firms in our two example supply networks obtained according to the old notion, (i) when edges are directed from supplier to buyer (showing the direction of material and service flows), and (ii) under the reverse interpretation (showing the direction of payment flows from buyers to sellers).  
%While levels are simply reflected about some value
%the same under reversal
%for our new notion, 
%(apart from an overall shift in level), 
It is apparent that with the old notion there is a big change in levels, the relevance of which is unclear.  Unless there is a good reason to favour basal nodes, we propose that our symmetric notion is better.

There have been some other approaches to rectifying the limitations of the original notion of trophic level.  Dominguez et al \cite{D+} obtain a `basal set' of nodes and eliminate all edges within that set.  Moutsinas et al \cite{Moutsinas} define levels using a pseudo-inverse of $L$.  These solutions allow application to networks without basal nodes but they don't possess symmetry with respect to reversal of edge directions nor a natural notion of maximal incoherence.  Another way to quantify trophic incoherence is to find the smallest number of edges to delete to obtain an acyclic graph \cite{T}, but it has some defects \cite{LBL}.   The smallest number is called the ``agony'' of the network.  Our trophic analysis provides a useful upper bound for agony, given by the number of edges with negative height difference, and could provide a useful heuristic for its exact computation.

In a very recent paper, \cite{KIII} decompose flows on a network into the sum of a potential part and a circulating part.  This looks a very nice approach, though it requires specifying conductivities for each edge as well as the flow on it, instead of specifying a target height difference for each edge.  The analysis has strong connections with ours, in particular the minimisation principle to determine the potential and an electrical interpretation (see SI).  Further work is required to make comparisons.

\section{Robustness of local computation}
If we determine trophic levels on a piece of a network by truncating the network at some distance from a chosen node, measured for example by the number of edges in either direction, how robust is the outcome to the truncation?

First we take care of the arbitrariness of the zero of trophic levels.  The simplest way to do that is to take the chosen node to be always at height zero.

Next, we refine the question because the trophic levels near the boundary of the piece of the network may change significantly with the truncation.  We ask how much the trophic levels change on a connected subset of the network containing the chosen node, which we will call zone 1, given a buffer zone 2 chosen so that there are no direct edges in either direction between zone 1 and the outside, called zone 3.  We choose the buffer zone so that in addition the union of zones 1 and 2 is connected (the only way this can not be satisfied is if zone 2 contains nodes which are not connected to zone 1 by a path in zone 2, in which case one can just throw them out).

\begin{figure}[htbp] %  figure placement: here, top, bottom, or page
   \centering
   \includegraphics[width=8.4cm]{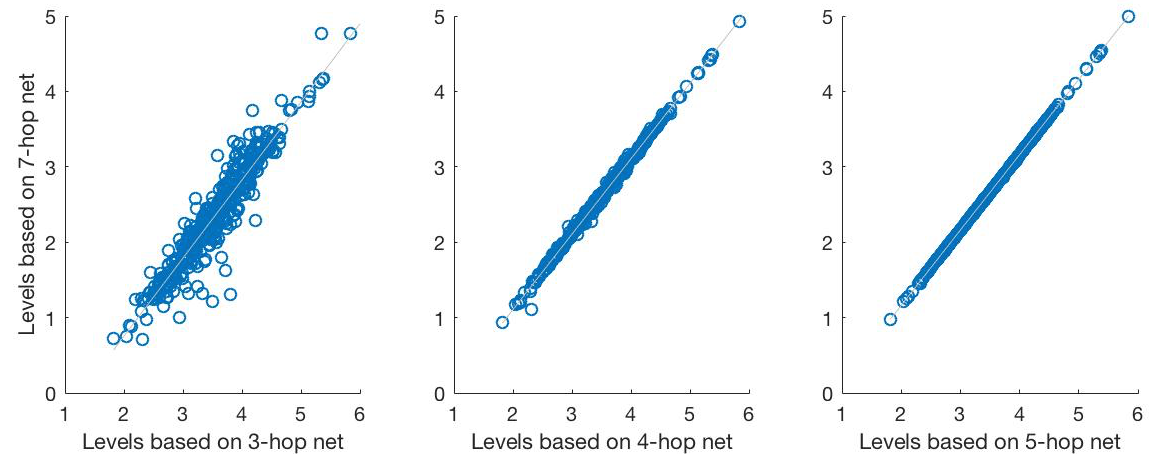} 
   \caption{Plot of the new trophic levels of buyers and suppliers in a 2-hop neighbourhood of General Motors calculated on networks constructed by sampling neighbourhoods of increasing size (3,4 and 5-hops) (horizontal axes), versus the levels of the same set of nodes calculated on a larger 7-hop neighbourhood (vertical axes). Supply chain information compiled from Bloomberg L.P.}
   \label{fig:GM}
\end{figure}

Figure~\ref{fig:GM} shows the outcome of a test, taking zone 1 to be the set of suppliers and buyers of General Motors (GM) 2 hops from GM, and computing the effects on the trophic levels in zone 1 of truncation of the network at 3, 4 and 5-hops respectively (i.e.~allowing a zone 2 buffer), compared to truncating at 7-hops. One can see that the trophic levels on zone 1 stabilise quite rapidly.
% two hops of GM, zone 2 to be those at 3 hops from GM, and computing the effects on the trophic levels in zone 1 of truncation of the network at  3 or 7 hops.  One can see that the trophic levels on zone 1 stabilise quite rapidly.

In the SI, we give some theoretical analysis to support the general conclusion that the levels on zone 1 are robust to changes on zone 3.

\section{Connections to other network properties}
A large part of the interest of the original notion of trophic coherence was its relation to network properties such as the stability of equilibria of Lotka-Volterra dynamics on the network \cite{JDDM}, the dynamics of spreading processes \cite{KJ1}, prevalence of cycles \cite{JJ}, other motifs \cite{KJ2}, intervality \cite{D+} and normality \cite{J}.  We show here that the new notion of trophic coherence has similar connections, even stronger, and it enlarges the scope of application because it does not require basal nodes.  We examine three of the properties.

\subsection{Normality}
A directed network is said to be {\em normal} if its weight matrix $W$ commutes with its transpose $W^T$:
\begin{equation}
W W^T = W^T W.
\end{equation}
Note that $W^T$ represents the same weighted network but with all the edges reversed.  Empirical directed networks are often highly non-normal \cite{A+}.  The term ``normal'' came from people who spent their lives with self-adjoint operators and unitary operators, both of which are normal, but people working in stability of ordinary differential equations are fully cognizant that most matrices are not normal.

For the unweighted case of an adjacency matrix $A$, normality implies the imbalance vector $v=0$.  This is because $(A^TA)_{mn}$ is the number of sources in common to nodes $m$ and $n$, and $(AA^T)_{mn}$ is the number of sinks in common.  In particular, $(A^TA)_{nn} = w_n^{in}$ and $(AA^T)_{nn}=w^{out}_n$, so $A^TA=AA^T$ implies that $w^{in}=w^{out}$ and $v=0$.

When $v=0$ we say a network is {\em balanced}.  A network is balanced if and only if its trophic incoherence $F_0 = 1$ (see SI).  
So normal unweighted networks are maximally incoherent.

Another special case of normality is symmetric networks $W = W^T$.  If $W$ is symmetric then the imbalance vector $v=0$.  So symmetry implies maximal incoherence. 

The concept of normality is broader than either of these, however.  Normality of $W$ is equivalent to existence of a unitary matrix $U$ such that $U^*WU$ is diagonal \cite{TE} (a {\em unitary} matrix is a complex-valued matrix $U$ such that $U^*U=I$, where $U^*$ is the complex conjugate of the transpose of $U$).  The diagonal elements of $U^*WU$ are the eigenvalues $\lambda_j$ of $W$ (with multiplicity).  From this we obtain the following extension of the result for symmetric networks:~if $W$ is normal and has all eigenvalues real then $F_0=1$ (see SI).  Perhaps the restriction to real eigenvalues is not necessary but we did not succeed in proving that.

Maximal incoherence, however, is not equivalent to normality.  There are non-normal networks with $v=0$ and hence maximal incoherence, e.g.
\begin{equation}
W = \left[\begin{array}{ccc} 1 & 1 & 0 \\ 0 & 0 & 1 \\ 1 & 0 & 0 \end{array} \right].
\label{eq:motif_3}
\end{equation}
%We suspect there are also unbalanced networks which are normal [MAKE AN EXAMPLE].

\begin{figure}[htbp] %  figure placement: here, top, bottom, or page
   \centering
   \includegraphics[width=8.4cm]{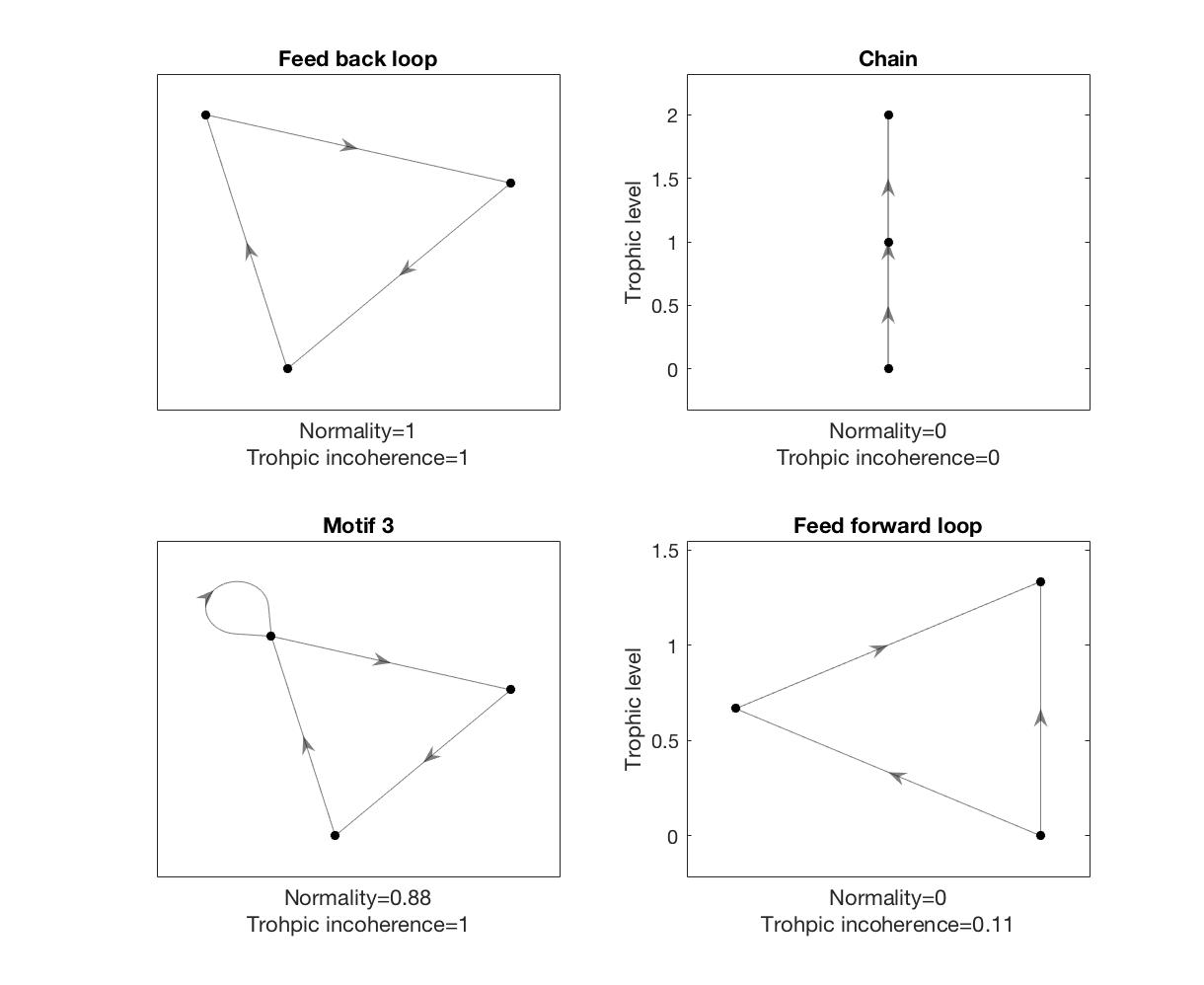} 
   \caption{Four simple motifs illustrating relationship between incoherence and normality: for the unweighted case of an adjacency matrix $A$, normality ($\nu=1$) implies the imbalance vector $v=0$, thus $F_0 = 1$. This is illustrated by the feed-back loop (top left). However maximal incoherence is not equivalent to normality - motif 3 (bottom left) demonstrates one can have $F_0 = 1$ without $\nu=1$ (here motif 3 [\ref{eq:motif_3}] is non-normal ($\nu=0.88$)). If $W$ is maximally coherent ($F_0=0$) then all its eigenvalues are 0, so $\nu=0$ and it is maximally non-normal. This is illustrated by the chain (top right). However one can have $\nu=0$ without $F_0=0$. This is demonstrated by the feed-forward motif (bottom right), which has $\nu=0$ but $F_0=0.11$.}
   \label{fig:motifs}
\end{figure}

Nevertheless, the extent to which a network is normal seems to be positively correlated with its trophic incoherence $F_0$.  The degree of normality of a network can be quantified by
\begin{equation}
\nu = \frac{\sum_j |\lambda_j|^2}{\| W \|_F^2},
\label{eq_nu}
\end{equation}
where $\| W \|_F = \sqrt{\sum_{mn} |w_{mn}|^2}$ is called the Frobenius norm of $W$, and $\lambda_j \in \C$ are the eigenvalues of $w$ (with multiplicity).  The literature uses $\sqrt{\|W\|_F^2-\sum_j |\lambda_j|^2}$ as a quantifier of non-normality, but we consider it simpler to use $\nu$.  The normality $\nu$ of $W$ lies in the interval $[0,1]$, with $\nu=1$ if and only if $W$ is normal \cite{TE}.  
If $W$ is maximally coherent ($F_0=0$) then all its eigenvalues are 0 (SI), so $\nu=0$ and it is maximally non-normal.  
But one can have $\nu=0$ without $F_0=0$, for example the feed-forward motif (see Figure~\ref{fig:motifs}) with
\begin{equation}
W = \left[ \begin{array}{ccc} 0 & 1 & 1 \\ 0 & 0 & 1 \\ 0 & 0 & 0 \end{array} \right],
\label{eq:ff}
\end{equation}
for which $h = [-2/3\ 0\ 2/3]^T$ and $F_0=1/9$.

\begin{figure}[htbp] %  figure placement: here, top, bottom, or page
   \centering
   \includegraphics[width=8.5cm]{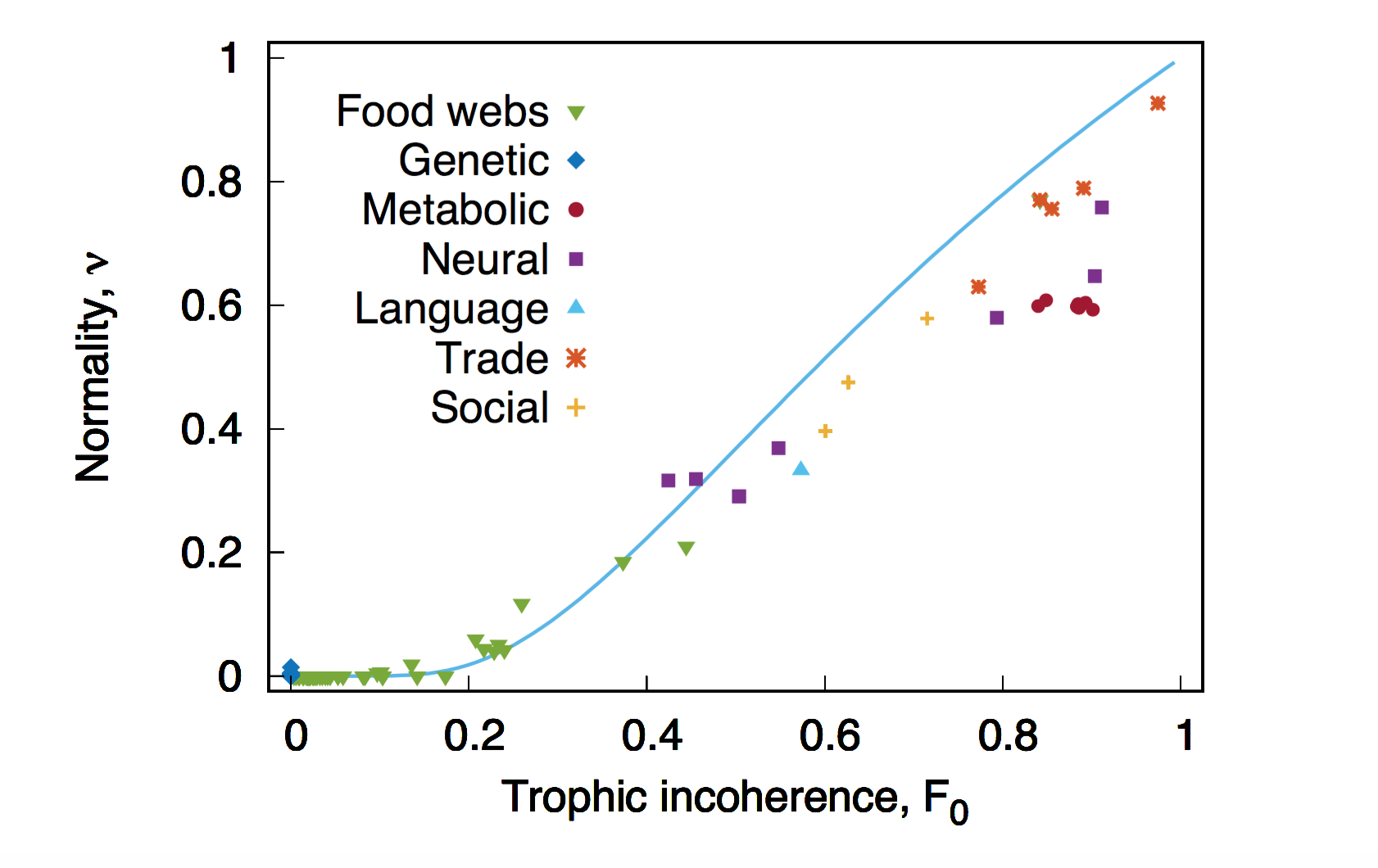}%{NuvZ} 
   \caption{Normality $\nu$ against trophic incoherence $F_0$
   %coherence $\bar{z}$
   for some networks. The curve corresponds to the coherence-ensemble expectation $\overline{\nu}=\exp(1-1/F_0)$.}
   \label{fig:nu}
\end{figure}
Figure~\ref{fig:nu} shows normality against trophic incoherence for some real networks.
We see that normality increases with $F_0$, but not linearly.
In the SI we present heuristic arguments in favour of a  relationship between them of the form $\nu \approx \exp (1-1/F_0)$.
This is consistent with a relationship between normality and the old notion of trophic coherence \cite{J}.

\subsection{Stability}
Next we discuss how dynamical processes on networks are affected by their trophic coherence.

A simple dynamical model for contagion on a weighted network in discrete time is
\begin{equation}
x'_n = \sum_m x_m w_{mn} / r,
\end{equation}
where $x_n \ge 0$ represents the amount of infection at node $n$ at some time, $x'_n$ the amount at the subsequent time, and $r>0$ is a reduction factor.  We wish to know whether the total infection $\|x\|_1 = \sum_n x_n$ on the network will grow or decay.  In vector-matrix form the solution after time $t \in \Z_+$ is
\begin{equation}
x(t) = x(0) W^t /r^t .
\end{equation}
The answer (see SI) is that if $\rho < r$ then $\|x(t)\|_1 \to 0$ as $t \to \infty$, whereas if $\rho > r$ and {\em condition $K$}: $x_n>0$ for some node $n$ in or leading to a ``key'' communicating class -- then $\|x(t)\|_1 \to \infty$, where the {\em spectral radius} $\rho$ of $W$ is the largest absolute value of the eigenvalues of $W$.  
Actually, because $W$ has all entries non-negative, it has a real positive eigenvalue of maximum modulus, so that is $\rho$.
Indeed, under condition $K$, 
\begin{equation}
t^{-1} \log \|x(t)\|_1 \to \log(\rho/r) \mbox{ as } t \to \infty.
\label{eq:rho}
\end{equation}

We have already mentioned that a maximally coherent network has all its eigenvalues 0, so $F_0=0$ implies $\rho=0$.  This suggests that $\rho$, scaled by a suitable measure of the strength of $W$, might correlate positively with $F_0$.  The strength of $W$ can be measured by any norm, for example the 2-norm $\|W\|_2$.  This can be defined in various ways, of which perhaps the simplest is that $\|W\|_2^2$ is the largest eigenvalue of $W^TW$ (which is necessarily real and non-negative and is equal to that for $WW^T$).
For any operator-norm, $\rho \le \|W\|$.  Thus $\rho/\|W\|$ is contained in $[0,1]$, like $F_0$.  An advantage of the particular choice of the 2-norm is that $\rho = \|W\|_2$ if $W$ is normal.  So we define the {\em scaled spectral radius}
\begin{equation}
    \rho_s = \rho/\|W\|_2
\end{equation}
Then we deduce from the subsection on normality various cases with simultaneously $F_0=1$ and $\rho_s= 1$.

\begin{figure}[htbp] %  figure placement: here, top, bottom, or page
   \centering
   \includegraphics[width=8.4cm]{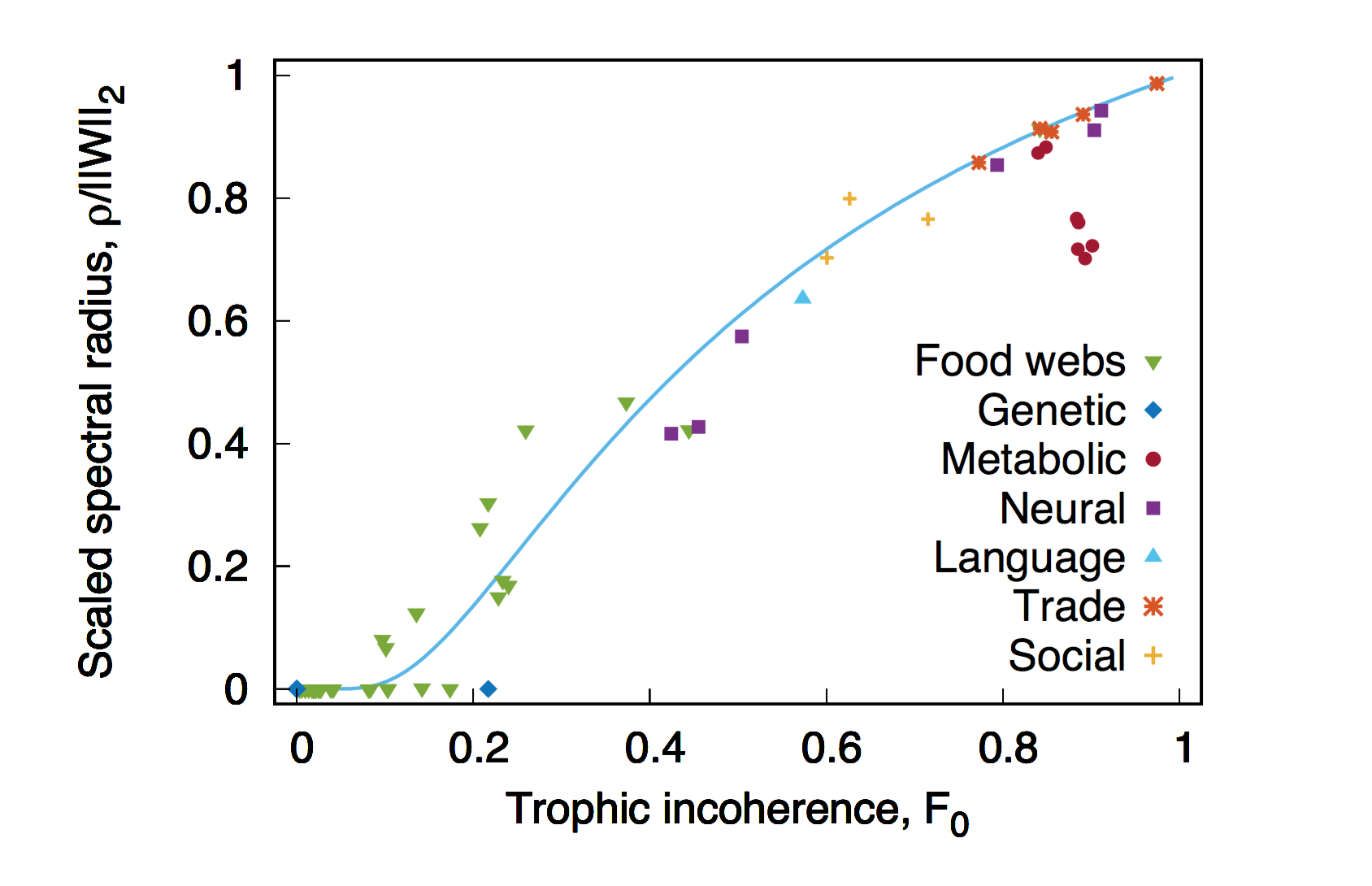} 
   \caption{Scaled spectral radius $\rho/\|W\|_2$ against trophic incoherence $F_0$ for some networks.  The curve corresponds to the coherence-ensemble prediction of $\rho_s = \exp(\frac12 (1-1/F_0))$.}
   \label{fig:specrad}
\end{figure}

Thus we look at how $F_0$ correlates with the scaled spectral radius $\rho_s$ in Figure~\ref{fig:specrad}.
In the SI we give heuristic arguments in favour of a relation $\rho_s \approx \exp(\frac12(1-1/F_0))$.

We can also consider a simple dynamical model for contagion in continuous time:
\begin{equation}
\dot{x}_n = \sum_m x_m w_{mn} - r x_n,
\end{equation}
with $r$ a recovery rate.  The solution can be written in vector-matrix form as
\begin{equation}
x(t) = x(0) e^{(W-rI)t}.
\end{equation}
Again one can ask whether the total infection $\|x(t)\|_1$ grows or decays.  This is now a question of the maximal real part of the eigenvalues of $W$, but because $W$ is non-negative, the maximal real part of eigenvalues is actually $\rho$.  So the answer is growth for $\rho>r$, decay for $\rho<r$.  So again it is interesting to link $\rho$ with $F_0$.

Some other dynamics on networks is discussed in the SI.

\subsection{Cycles}
A {\em cycle} in a directed network is a closed walk in it.  In contrast to some of the literature, we allow repeated edges and repeated nodes.  In particular, we allow a cycle to be a periodic repetition of a shorter cycle.  The {\em weight} $w_\gamma$ of a cycle $\gamma$ is the product of the weights along its edges.

A maximally coherent network ($F_0=0$) has no cycles, because it has height difference $+1$ for every edge whereas along a cycle the nett change in height has to be zero.  There are acyclic graphs with $F_0>0$, however, for example the feedforward motif [\ref{eq:ff}].

A maximally incoherent network ($F_0=1$) must have cycles.  This is because it is balanced and so some of the flow that leaves a node must eventually come back to it (see SI).  In fact, we deduce that every edge is in at least one cycle.

So these results suggest some relation between trophic incoherence $F_0$ and a quantifier of cyclicity.

The total weight of cycles of length $p$ is given by the trace of the $p^{th}$ power of $W$: $\tr\ W^p$, because $(W^p)_{mn} = \sum_j w_{n_0n_1}\ldots w_{n_{p-1}n_p}$ and the trace of a matrix is the sum of its diagonal entries.  One might expect it to behave asymptotically exponentially as $p \to \infty$, but for example if $k$ points in a circle are each connected to just their clockwise neighbour by an edge of weight $x$, then $\tr\ W^p = kx^{p}$ when $p$ is a multiple of $k$, $0$ otherwise.  The tidy way to study the sequence $\tr\ W^p$ is to form the {\em zeta function} 
\begin{equation}
\zeta(z) = \exp\sum_{p=1}^\infty \frac{z^p}{p}\tr\ W^p 
\end{equation}
for complex $z$ close enough to $0$  (some authors define $\zeta(z)$ to be the reciprocal of this). Then a notion of the cyclicity of $W$ is the reciprocal of the radius of convergence of the power series.  This is just $\limsup_{p \to \infty} (\tr\ W^p)^{1/p}$.  Using $\log\det = \tr \log$, the zeta-function can equivalently be written as $\det (I-zW)^{-1}$.  The reciprocal of its radius of convergence is the spectral radius $\rho$.  So actually, the appropriate measure of cyclicity is $\rho$ relative to some measure of the size of $w$.  We take again $\|W\|_2$ for the latter.  Thus cyclicity $\rho/\|W\|_2 = \rho_s$ is related to $F_0$ exactly as is the stability of our simple contagion processes.  In particular it is 1 for any normal network.

In the SI, we relate $\zeta$ to the {\em prime} cycles, those which are not repetitions of a shorter cycle, and furthermore to the {\em elementary} cycles, those which do not repeat a node.

\section{Discussion}
We have improved the definitions of trophic level and incoherence so that they can be applied to any directed network, not just those with basal nodes, to remove a bias from basal nodes, to make incoherence have a natural range from perfect coherence to maximal incoherence, and to make it possible to compute them locally in a network without having to compute them for the whole network.

We anticipate our improved notions being useful in many domains, from ecology, gene expression, neuroscience, supply networks and financial networks to linguistics and social networks.  The scope is enormous.

%Various refinements merit further study.  In particular, theory for ensembles of weighted networks, connection with \cite{KIII}... 

A further issue to address is the effect of splitting or merging nodes.  It may be appropriate to develop a  refined notion of trophic level to allow the target height differences between nodes to be specified rather than all being taken $+1$.

\matmethods{
%Please describe your materials and methods here. This can be more than one paragraph, and may contain subsections and equations as required. Authors should include a statement in the methods section describing how readers will be able to access the data in the paper.
\vspace{-0.5cm}
\subsection*{Mathematical Analysis}
Proofs of the mathematical results presented here are included in the SI.

\subsection*{Data Sources}
The data used to produce Figure~\ref{fig:Ythan} was downloaded from \cite{CE} and can be accessed from: \url{https://datadryad.org/stash/dataset/doi:10.5061/dryad.1mv20r6}. The data used to produce Figures~\ref{fig:IO} and \ref{fig:sectors} are from the OECD Input-Output Tables described and available here: \url{http://www.oecd.org/sti/ind/input-outputtables.htm}, from the OECD website. The data for Figure~\ref{fig:regulatory} was downloaded from \cite{Jothi} here: \url{https://www.ncbi.nlm.nih.gov/pmc/articles/PMC2736650/}. The data for Figure~\ref{fig:books} published in \cite{RG} was downloaded from \url{http://language.media.mit.edu/data}. The supply network data-sets presented in Figures~\ref{fig:supply}-\ref{fig:comparison} and used for the analysis presented in Figure~\ref{fig:GM} are based on supply-chain relationships compiled from Bloomberg L.P. supply chain function. Bloomberg's database compiles information from a wide variety of sources to provide a view of global supply chains at the firm level. More information on this data can be obtained from Bloomberg L.P. or \cite{D}. To construct our networks of supplier–buyer relationships, starting from a focal firm of interest we then followed links identified by the Bloomberg database. These data-sets could with Bloomberg's permission be made available on request, or re-compiled from Bloomberg.
The data used for Figures~\ref{fig:nu} and \ref{fig:specrad} can be downloaded from \url{https://www.samuel-johnson.org/data}, along with a list with references to the original sources.
All code used to make empirical and computational analysis of public data and data-files is  available for download at this \href{https://github.com/BazilSansom/How-directed-is-a-directed-network.git}{Github repository}, where we also provide a Matlab toolbox for the easy implementation of the methods we have introduced and related analysis.
% [CE] Cirtwill, Alyssa R.; Eklöf, Anna (2018), Data from: Feeding environment and other traits shape species' roles in marine food webs, Dryad, Dataset, https://doi.org/10.5061/dryad.1mv20r6
%[D] Davenport, Richard. “Bloomberg’s Supply Chain Algorithm: Providing Insight into Company Relationships. Bloomberg Terminal. 18 Aug 2012
}

\showmatmethods{} % Display the Materials and Methods section

\acknow{The support of the Economic and Social Research Council (UK), through grant ES/R00787X/1, is gratefully acknowledged.  This was awarded via a call from the Instability hub of the Rebuilding Macroeconomics programme of the National Institute for Economic and Social Research (NIESR).  We are grateful to other members of our project team for their comments, especially Nicholas Beale for asking for a quantification of coherence that 
does not depend on having basal nodes and has
a clear maximum and minimum,
and to other members of the Instability hub for their comments and interest.  We are also grateful to Giannis Moutsinas for sharing their approach to the subject, and to Mark Pollicott for useful discussion about the zeta function.
SJ and RSM acknowledge support of the Alan Turing Institute under 
EPSRC grant EP/N510129/1
and
Fellowship grant
TU/B/000101.
%This work was supported by The Alan Turing Institute under the EPSRC grant EP/N510129/1
}

\showacknow{} % Display the acknowledgments section

% Bibliography
%\bibliography{pnas-sample}
%\subsection*{References}

%\newpage
\section*{Supporting Information}

\subsection*{Solutions of $\Lambda h = v$}
The graph Laplacian $\Lambda$ is not invertible:~for any constant vector $h$, $\Lambda h = 0$.  Indeed for any $h$ that is constant on connected components of the network, $\Lambda h=0$, and the kernel of $\Lambda$ is precisely this set of $h$.  Similarly, for any $h$, the components of $\Lambda h$ on each connected component of the network add up to zero, and this property characterises the range of $\Lambda$.  Now the imbalance vector $v$ has the special property that the sum of its components over any connected component of the network is zero.  Thus it follows that $\Lambda h = v$ always has a solution $h$, and the general solution is given by adding any vector that is constant on each connected component.

\subsection*{Range for trophic incoherence $F_{0}$}
Here we prove that $0\le F_{0} \le 1$ with $F_{0}=0$ iff all height differences $z_{mn} = 1$ and $F_{0}=1$ iff all height differences are $0$.

First, we explain that the trophic heights $h$ solving $\Lambda h = v$ correspond to the minima of the trophic confusion function
\begin{equation}
F(h) = \frac{\sum_{mn} w_{mn}(h_n-h_m-1)^2}{\sum_{mn} w_{mn}}
\end{equation}
over all possible assignments of heights $h_n, n \in N$.  This is because the second derivative of $F$ is positive semi-definite, so all critical points are minima, and by differentiating with respect to each $h_n$, the equation for critical points is $\Lambda h = v$.  Furthermore, the minimum value of this expression is $F_{0}$.

Since $F(h)\ge 0$ for all $h$, we see that $F_{0}\ge 0$.  Furthermore, $F_{0}=0$ iff all height differences are $1$.
Next, putting all heights equal, say to 0, denoted by $\bf{0}$, gives $F({\bf 0}) = 1$, so $F_{0} \le 1$.  Now if $F_{0}=1$ at some $h$ then because $F({\bf 0})=1$ and the second derivative of $F$ is positive semi-definite with null space given by constants on each connected component, then $h-\bf{0}$ must be in this nullspace, i.e.~$h$ is constant on each connected component.  Thus all height differences along edges are zero.

\subsection*{Mean height difference}
The mean height difference 
\begin{equation}
    \bar{z} = \frac{\sum_{mn} w_{mn}(h_n-h_m)}{\sum_{mn} w_{mn}}
\end{equation} 
is $1-F_0$.  To prove this, write the trophic confusion function as
\begin{equation}
    F(h) = \sigma^2 + \bar{z}^2-2\bar{z}+1,
\end{equation}
with 
\begin{equation}
    \sigma^2 = \frac{\sum_{mn} w_{mn}(h_n-h_m-\bar{z})^2}{\sum_{mn} w_{mn}}.
\end{equation}
If $h$ minimises $F$ then $F(\alpha h)$ must be minimised over $\alpha \in \R$ at $\alpha=1$.  But
\begin{equation}
F(\alpha h) = \alpha^2 (\sigma^2+\bar{z}^2) - 2\alpha \bar{z} + 1,
\end{equation}
which has unique minimum at $\alpha = \frac{\bar{z}}{\sigma^2+\bar{z}^2}$ (unless $\sigma=\bar{z}=0$).  Thus $\bar{z} = \sigma^2 + \bar{z}^2$.  But $\sigma^2 + \bar{z}^2 - 2\bar{z}+1 = F_0$.  So $\bar{z} = F_0+2\bar{z}-1$, hence $\bar{z}=1-F_0$.  If $\sigma=\bar{z}=0$, we see that $F_0=1$ and hence $\bar{z}=1-F_0$ is satisfied in that case too.

\subsection*{Electrical interpretation}
Our new notion of trophic levels can be given an electrical interpretation.  The edge weights are conductivities of bidirectional connectors between nodes.  Current $v_n$ is injected into (or extracted from, according to sign) each node $n$.  The resulting voltages (modulo an arbitrary overall shift) are the trophic heights $h_n$.  One could imagine the currents $v_n$ as being generated by making a copy of all the incoming and outgoing edges of node $n$ and imposing a voltage difference of $+1$ on all its input nodes and $-1$ on all its output nodes, relative to $n$.

\subsection*{Robustness of trophic levels to truncation of the network}
We recall that we choose a connected subset called zone 1 and fix the height of one of its nodes (or the weighted average of its nodes) to be $0$.  We choose a buffer zone 2 so that there are no direct connections between zone 1 and the outside, called zone 3, and so that the union of zones 1 and 2 is connected.  

Then the equation $\Lambda h = v$ for the heights can be broken into the block form
\begin{eqnarray}
\Lambda_{11}h_1 + \Lambda_{12} h_2 \quad\quad\quad\quad &=& v_1 \\
\Lambda_{21}h_1 + \Lambda_{22} h_2 + \Lambda_{23}h_3 &=& v_2 \\
\Lambda_{32}h_2 + \Lambda_{33} h_3 &=& v_3.
\end{eqnarray}
Changes to the outside zone 3 can affect $v_2$ and the diagonal part of $\Lambda_{22}$.  Let us suppose that the total weights of connections in each direction between zone 3 and each node of 2 are given.  Thus $v_2$ and $\Lambda_{22}$ are fixed.
Let $\bar{h}$ be the solution for the reference case where all of zone 3 is amalgamated to a single node.  
By the connectedness assumption, $\bar{h}$ exists and is unique up to an overall shift.
Let $\tilde{h} = h-\bar{h}$ with $h$ the solution for the true zone 3, subtracting the single number $\bar{h}_3$ from each element of $h_3$.  Then
\begin{eqnarray}
\Lambda_{11} \tilde{h}_1 + \Lambda_{12}\tilde{h}_2 \quad\quad\quad\quad &=& 0 \label{eq:11} \\
\Lambda_{21} \tilde{h}_1 + \Lambda_{22}\tilde{h}_2 + \Lambda_{23}\tilde{h}_3 &=& 0 .
\end{eqnarray}
By the connectedness of zone 1, $\Lambda_{11}$ is invertible modulo overall shifts, on the subspace such that the sum of the components is zero.  We have taken care of overall shifts by fixing a node of zone 1 to be at height 0.  The sum of the components of $\Lambda_{12}\tilde{h}_2$ is automatically zero, because taking the sum of (\ref{eq:11}) over components in zone 1, $\Lambda_{11}\tilde{h}_1$ gives $0$.
So
\begin{equation}
\tilde{h}_1 = -\Lambda_{11}^{-1}\Lambda_{12}\tilde{h}_2.
\end{equation}
Similarly, by connectedness of the union of zones 1 and 2, and substituting the above,
\begin{equation}
\tilde{h}_2 = -(\Lambda_{22}-\Lambda_{21}\Lambda_{11}^{-1}\Lambda_{12})^{-1}\Lambda_{23}\tilde{h}_3
\end{equation}
Thus the desired answer is
\begin{equation}
\tilde{h}_1 = \Lambda_{11}^{-1}\Lambda_{12}(\Lambda_{22}-\Lambda_{21}\Lambda_{11}^{-1}\Lambda_{12})^{-1}\Lambda_{23}\tilde{h}_3
\end{equation}
Thus by taking norms throughout (for example the weighted sum $\|h\| = \sum_n u_n |h_n|$ and the corresponding operator norm), we obtain a bound on the changes to the levels on zone 1 in terms of a bound on the changes to the levels on the part of zone 3 connecting directly to zone 2:
\begin{equation}
\|\tilde{h}\| \le \|\Lambda_{11}^{-1}\| \|\Lambda_{12}\| \|(\Lambda_{22}-\Lambda_{21}\Lambda_{11}^{-1}\Lambda_{12})^{-1}\| \|\Lambda_{23}\tilde{h}_3\|.
\end{equation}
The latter is unknown in general, but the formula gives some idea of how much the levels change on zone 1 on incorporating more detail about zone 3.  In particular, if zone 1 is well connected in the sense that $\|\Lambda_{11}^{-1}\|$ is not large, and zones 1 and 2 are well connected in the sense that $\|(\Lambda_{22}-\Lambda_{21}\Lambda_{11}^{-1}\Lambda_{12})^{-1}\|$ is not large then $\tilde{h}_1$ is not very sensitive to changes $\tilde{h}_3$ to the levels in zone 3.

An alternative to fixing the height of a node in zone 1 is to consider height vectors as equivalent if they differ by an overall shift and use a norm that pays attention only to height differences, e.g.~$\|h\| = \sum_{mn} w_{mn} |h_n-h_m|$.

\subsection*{Balanced iff maximally incoherent}
If $v=0$ then $\Lambda h = 0$ so $h$ is constant on connected components, so $F_0=1$.  Conversely, if $F_0=1$ then $h$ is constant on connected components, so $v=\Lambda h = 0$.

Note that it follows that maximally incoherent networks have no basal nodes (more precisely, any basal node is connected to no other nodes).

\subsection*{Normal with all eigenvalues real implies maximally incoherent}
A normal matrix $W$ can be written as $U\lambda U^*$ for some unitary matrix $U$, where $\lambda$ is the diagonal matrix of the eigenvalues $\lambda_j$ of $W$ (repeated according to multiplicity).  Then
(using $\bar{z}$ for the complex conjugate of $z$)
\begin{eqnarray}
w_n^{in} &=& \sum_m w_{mn} = \sum_{m,j} U_{mj}\lambda_j \bar{U}_{nj} \\
w_n^{out} &=& \sum_m w_{nm} = \sum_{m,j} U_{nj}\lambda_j \bar{U}_{mj}.
\end{eqnarray}
But $W$ and hence $w^{out}$ is real so we can take the complex conjugate of the second equation and deduce that
\begin{equation}
v_n = w_n^{in}-w_n^{out} = \sum_{m,j} U_{mj} (\lambda_j-\bar{\lambda}_j) \bar{U}_{nj}.
\end{equation}
From this we see that if all the eigenvalues are real then $v=0$.  Then from the preceding item, $F_0=1$.

\subsection*{Maximal coherence implies normality zero}
If $W$ is maximally coherent then the level difference for each edge is $+1$ so, arranging the nodes in order of height, the matrix $W$ is upper triangular with zero diagonal.  It follows that all its eigenvalues are 0.  Hence $\nu=0$.

\subsection*{Stability of contagion processes}
For $x(t) = x(0) W^t/r^t$, we have 
\begin{equation}
\|x(t)\|_1 \le \|x(0)\|_1 \|W^t\|_1 /r^t,
\end{equation}
using the induced operator-norm on $W$, so
\begin{equation}
t^{-1} \log \|x(t)\|_1 \le t^{-1} \log \|x(0)\|_1 + t^{-1} \log \|W^t\|_1 - \log r.
\label{eq:bound}
\end{equation}
But for any operator-norm on $W$, $t^{-1} \log \|W^t\| \to \log \rho$ as $t\to \infty$ \cite{RS}.  So if $\rho < r$ then $\|x(t)\|_1 \to 0$ as $t\to \infty$.

In the other direction, we need theory for non-negative matrices $W$, e.g.~\cite{BP}.  A node in a directed graph is {\em recurrent} if there is a cycle through it.  Two recurrent nodes {\em communicate} if there is a cycle through both.  The set of recurrent nodes can be decomposed into {\em communicating classes}, subsets in which each pair of nodes communicate and between which no pair of nodes communicate.  The eigenvalues of $W$ consist of the eigenvalues of its restrictions $W_c$ to each communicating class $c$ and an eigenvalue 0 for each non-recurrent node.  The {\em period} $P$ of a communicating class $c$ is the highest common factor of the lengths of all cycles in it.  The communicating class $c$ can be decomposed into $P$ {\em cyclic classes}, whose nodes can only be reached from each other in a multiple of $P$ steps.  They can be labelled $c_0,\ldots c_{P-1}$ so that one can get from $c_j$ to $c_k$ only in a number of steps congruent to $k-j$ modulo $P$.

On each cyclic class $c_j$, the restriction of $w^{P}$ is irreducible and aperiodic.  So by Perron-Frobenius theory \cite{BP} it has a simple positive eigenvalue $\lambda_1$ with positive eigenvector, and the remaining eigenvalues satisfy $|\lambda_k| < \lambda_1$.  Throughout this item, we consider left eigenvectors because we are interested in the action of $W$ on row-vectors $x$. The eigenvalues of $W_{c_j}^{P}$ on the cyclic classes of $c$ are related as follows.  If $xW_c^{P} = \lambda x$ with $x$ supported on $c_0$ and non-zero, then $xW_c$ is supported on $c_1$ and $(xW_c)W_c^P = xW_c^P W_c = \lambda xW_c$, so either $xW_c$ is an eigenvector for $W_c^P$ on $c_1$ with the same eigenvalue or it is zero.  If $xW_c=0$ then $\lambda x = xW_c W_c^{P-1} = 0$ so $\lambda=0$.  Thus $W_{c_j}^P$ have the same eigenvalues apart from possible 0s.  If the cyclic classes have different sizes, eigenvalues 0 must occur for all but the smallest ones.

From the non-zero eigenvalues $\lambda$ of $W_{c_j}^P$ we deduce that the non-zero eigenvalues of $W_c$ are the (complex) $P^{th}$ roots of $\lambda$ as follows.  Take an eigenvector $x$ on $c_0$ for $\lambda \ne 0$.  Let $\zeta$ be any $P^{th}$ root of $\lambda$.  Then $[\zeta^Px, \zeta^{P-1}xW_c, \ldots \zeta x W_c^{P-1}]$ is an eigenvector of $W_c$ with eigenvalue $\zeta$, where the components in the vector are grouped according to the cyclic classes $c_0,\ldots c_{P-1}$.  So the eigenvalues of $W_c$ are the $P^{th}$ roots of the non-zero eigenvalues of $W_{c_0}^P$, augmented by 0s.  The eigenvectors of $W_c$ can be extended to eigenvectors of $W$ on the whole network with the same eigenvalue.

If $x(0)\ge 0 $ is positive on some node of a cyclic class $c_j$ of a communicating class $c$ then by Perron-Frobenius theory,
\begin{equation}
\lambda_1^{-n} x(0)W_{c_j}^{nP} \to C \hat{x} \mbox{ as } n \to \infty
\end{equation}
for some $C>0$, where $\hat{x}$ is the Perron-Frobenius eigenvector on $c_j$ and $\lambda_1$ its eigenvector.  Furthermore $\lambda_1^{-n} x(0)W_c^{nP+s} \to C \hat{x}W_c^s$.  Including the rest of the edges,  
\begin{equation}
\lim_{n\to \infty} t^{-1} \log\|x(t)\|_1 \ge P^{-1}\log\lambda_1.
\label{eq:loglambda}
\end{equation}
We say a communicating class is ``key'' if its $\lambda_1 = \rho^P$.  There is always at least one such.
Thus if $x(0)$ is positive on some node of a key communicating class, then, combining with [\ref{eq:bound}] and [\ref{eq:loglambda}],
\begin{equation}
    t^{-1}\log\|x(t)\|_1 \to \log(\rho/r) \mbox{ as } t\to \infty.
\end{equation}
Similarly, if $x(0)$ is positive on a node leading to a key communicating class then in finitely many steps $x(t)$ is positive on some node of that class and hence the same result follows.

\subsection*{Other dynamics}
The context in which trophic coherence was first proposed \cite{JDDM} is that of Lotka-Volterra dynamics for populations of species in an ecosystem.  This is somewhat difficult to treat because if $w$ quantifies how much one species eats of another this does not give a complete specification of the population dynamics.  But as in \cite{JDDM}, one can propose
\begin{equation}
\dot{x}_n = x_n \left(r_n - \sum_m w_{nm}x_m + \eta \sum_m x_m w_{mn}-\kappa_n x_n\right),
\end{equation} 
where $r_n$ is a natural birth or death rate (according to sign) for species $n$, the negative sum accounts for species $n$ being eaten, the positive sum accounts for the enhancement of population of species $n$ from what it eats, with an efficiency factor $\eta$, and the final term accounts for effects of intraspecies competition not included in cannibalism ($w_{nn}$).  Write it in the form
\begin{equation}
\frac{d}{dt} \log x = r -Bx,
\end{equation}
where $\log x$ stands for the vector with components $\log x_n$.

One first question is whether this has any positive equilibria.  The equilibria are given by choosing any subset of species to be extinct and the rest to satisfy $Bx = r$ where the rows and columns corresponding to extinct species have been deleted.  To be physical the remaining components of $x$ must all be positive.

Given a positive equilibrium $x$, possibly of a subsystem given by deleting extinct species, a second question is whether it is stable.  The linearised equations for deviations $\xi$ from an equilibrium are 
\begin{equation}
\dot{\xi}_n = -x_n\sum_m B_{mn}\xi_n
\end{equation}
So even if we know $B$, the linearised equations are not completely determined because we need to know the equilibrium $x$.

Similarly, economic dynamics can be proposed on supply networks \cite{MB} and the question arises whether there is a relation between stability and trophic coherence.

\subsection*{Ensemble relation of normality to incoherence}

It is possible to relate trophic coherence with various other topological features by considering 
ensembles of random graphs \cite{JJ}. 
The `coherence ensemble' is the set of all
unweighted, directed networks with given in- and out-degree sequences and given trophic coherence.
For example, using the standard definition of trophic incoherence $q$, the expected value of the 
spectral radius $\rho$ in the coherence ensemble is
\begin{equation}
 \overline{\rho}=e^\tau,
 \label{eq_rho}
\end{equation}
where
\begin{equation}
\tau=\ln \alpha + \frac{1}{2\hat{q}^2}-\frac{1}{2 q^2}
\label{eq_tau}
\end{equation}
(and we use a bar to represent coherence-ensemble expectation).
Here, $\hat{q}$ is the expected trophic incoherence in the `basal ensemble', 
and $\alpha=\langle w^{in} w^{out}\rangle/\langle w\rangle$ is the branching factor, but for current purposes 
we need not discuss these magnitudes in detail.
In previous work the trophic coherence was measured with the incoherence parameter $q$, which corresponds to the 
standard deviation over trophic differences when the average trophic difference is $1$.
Using the new definition of levels we are proposing here, the equivalent of this
magnitude is
% $\eta$, as given by Eq. (\ref{eq_eta}).
\begin{equation}
 \eta=\sqrt{\frac{F_0}{1-F_0}},
\label{eq_eta_SI}
 \end{equation}
as given in the main text.
% by Eq. (\ref{eq_eta}).
% It is only necessary to note that 

We note that the ratio between the expected spectral radius
for a given coherence,
$\overline{\rho}$, and the value corresponding to a maximally incoherent network,
\begin{equation}
 \overline{\rho}_{max}=\lim_{\eta\rightarrow \infty} e^{\tau},
\end{equation}
depends only on trophic coherence:
\begin{equation}
 \frac{\overline{\rho}}{\overline{\rho}_{max}}=\exp\left(-\frac{1}{2 \eta^2}\right).
 \label{eq:rhobar}
\end{equation}

In the main text we measure normality with
\begin{equation}
\nu = \frac{\sum_j |\lambda_j|^2}{\| W \|_F^2}.
\label{eq_nu_SI}
\end{equation}
A normal network (if unweighted) is, as described in the main text, a balanced network, which is maximally incoherent ($F_0=1$). In this case, we have $\nu=1$. On the other hand, the greatest deviation from normality is achieved when 
$|\lambda_j|=0$ for all $j$, which is the case of maximally coherent networks ($F_0=0$).
For networks in the coherence ensemble with $0\leq F_0 \leq 1$, we postulate that
\begin{equation}
 \frac{\sum_i|\lambda_i|^2}{\|W\|_F^2} \simeq \frac{\rho^2}{\rho_{max}^2},
\end{equation}
which amounts to assuming that the distribution of eigenvalues of $W$ within the spectral radius $\rho$ does not depend on trophic coherence.
This argument uses (i) $\|W\|_F^2 = \tr\ W^TW$, which in turn is the sum of the eigenvalues of $W^TW$, (ii) for $W$ normal the eigenvalues of $W^TW$ are precisely the squares of the absolute values of the eigenvalues of $W$, and (iii) normality is almost equivalent to maximal incoherence, as already discussed.
Combining this expression with Eqs.~[\ref{eq_eta_SI}], [\ref{eq:rhobar}] and [\ref{eq_nu_SI}], we have an approximate expression for the expected normality in the coherence ensemble:
\begin{equation}
 \overline{\nu}\simeq \exp\left(1-\frac{1}{F_0}\right).
 \label{eq_nu_bar_SI}
\end{equation}

Figure \ref{fig:nu} in the main text shows $\nu$ against $F_0$ for our set of empirical networks, alongside
Eq.~[\ref{eq_nu_bar_SI}]. The empirical values fall fairly close to the ensemble expectations, with high coherence 
corresponding to a maximal non-normality, and incoherence being associated with greater normality. In many cases the real 
networks are somewhat less normal than the ensemble prediction. This might be because these are relatively small networks in which
statistical fluctuations play a large role, and at intermediate values of trophic coherence there are more ways of being non-normal
than normal.

\subsection*{Ensemble relation with scaled spectral radius}
Using the results for the coherence ensemble again, in particular Eqns [\ref{eq:rhobar}] and [\ref{eq_eta_SI}],
we obtain that
\begin{equation}
    \rho_s = \frac{\rho}{\|W\|_2} \simeq \frac{\overline{\rho}}{\overline{\rho}_{max}} 
    =\exp\left( \frac12\left(1-\frac{1}{F_0}\right)\right).
\end{equation}
Here, no assumption on the distribution of the eigenvalues of $W$ is required, simply the fact that $\rho = \|W\|_2$ for maximally incoherent networks.

The fit in Figure~\ref{fig:specrad} is again reasonable.

\subsection*{Maximal incoherence implies cycles}
A maximally incoherent network is balanced.  Make a measure-preserving dynamical system in continuous time by converting each edge $mn$ to a tube of volume $V_{mn}>0$ of incompressible fluid with flow rate $w_{mn}$ from $m$ to $n$, splitting the resulting flow into $n$ in any way between the out-edges of $n$ consistent with their weights.  If none of the fluid originally in tube $mn$ comes back to that tube then after time $T$, tube $mn$ has ejected a volume $w_{mn}T$ of fluid that has to fit in the volume $\sum V_{jk}$ of the other tubes.  But that is finite, so for $T$ large enough we get a contradiction.  Hence there is a cycle through $mn$.  So each edge of a maximally incoherent network is on a cycle.

One could allow the nodes to have volume too.  The same argument works for infinite networks, by choosing the volumes to have a finite sum.

\subsection*{Zeta function}
The zeta function of the main text is a weighted version of the Bowen-Lanford zeta function (described in section 3.1 of \cite{Po}).  It can be related to the {\em prime} cycles, those which are not repetitions of a shorter cycle.  We consider two prime cycles to be the same if they differ only by a cyclic permutation.  We denote by $\mathcal{P}$ the set of prime cycles.  The formula is
\begin{equation}
\zeta(z) = \prod_{\gamma \in \mathcal{P}} (1-z^{|\gamma|} w_\gamma)^{-1},
\label{eq:zeta}
\end{equation}
where $|\gamma|$ is the length of $\gamma$ and $w_\gamma$ its weight.

Here is a proof of the identity (cf.~\cite{Po}).
\begin{align}
\nonumber
\log \prod_\gamma (1-z^{|\gamma|}w_\gamma)^{-1} = - \sum_\gamma \log (1-z^{|\gamma|}w_\gamma) =\\
\sum_\gamma \sum_{k\ge 1} \frac{(z^{|\gamma|}w_\gamma)^k}{k}
= \sum_{k\ge 1} \sum_\gamma |\gamma| \frac{z^{k|\gamma|}}{k|\gamma|} w_\gamma^k = \sum_{p\ge 1} \frac{z^p}{p} \tr\ W^p,
\end{align}
%\begin{figure*}
%\begin{eqnarray}
%\log \Pi_\gamma (1-z^{|\gamma|}w_\gamma)^{-1} &=& - \sum_\gamma \log (1-z^{|\gamma|}w_\gamma) = \sum_\gamma \sum_{k\ge 1} \frac{(z^{|\gamma|}w_\gamma)^k}{k} \\
%&=& \sum_{k\ge 1} \sum_\gamma |\gamma| \frac{z^{k|\gamma|}}{k|\gamma|} w_\gamma^k = \sum_{p\ge 1} \frac{z^p}{p} \tr\ W^p,
%\end{eqnarray}
%\end{figure*}
because every cycle is a repetition of some prime cycle $\gamma$, say $k$ times, its weight is $w_\gamma^k$ and there are $|\gamma|$ cyclic permutations of it.  The last expression is $\log \zeta(z)$, concluding the proof.

Eq.~[\ref{eq:zeta}] can be reduced to one in terms of ``elementary cycles'', those which do not repeat a node before closing.  They are prime and for a finite network there are only finitely many of them.  The formula is
\begin{equation}
    1/\zeta(z) = 1 + \sum_C \prod_{\gamma \in C} (-z^{|\gamma|}w_\gamma),
    \label{eq:cycleexp}
\end{equation}
where the sum is over non-empty collections $C$ of disjoint elementary cycles.
This provides a clean case of Cvitanovic's cycle expansion \cite{AAC}.

To prove [\ref{eq:cycleexp}], use $1/\zeta(z) = \det(I-zW)$ and the formula
\begin{equation}
    \det M = \sum_{\pi \in S_n} \epsilon_\pi M_{1\pi_1}\ldots M_{n\pi_n},
    \label{eq:det}
\end{equation}
for an $n\times n$ matrix $M$, where $S_n$ is the group of permutations of $\{1,\ldots n\}$ and $\epsilon_\pi$ is the sign of the permutation $\pi$ ($+1$ if $\pi$ can be written as an even number of transpositions, $-1$ for an odd number).  For $M=I-zW$, the only permutations for which the product in [\ref{eq:det}] is non-zero are those which can be written as a product of disjoint cyclic permutations corresponding to elementary cycles of period at least 2 in the network and the identity permutation on the remaining nodes.
The contribution of a collection $C_2$ (possibly empty) of disjoint elementary cycles of period at least 2 is
\begin{equation}
    \prod_{m \in C'}(1-zw_{mm}) \prod_{\gamma \in C_2} (-z^{|\gamma|}w_\gamma),
    \label{eq:C2}
\end{equation}
where $C'$ is the set of nodes not in $C_2$.  If there are no self-edges then $w_{mm}=0$ for all $m$ and there are no cycles of period 1, so adding in the case of the empty collection, we obtain [\ref{eq:cycleexp}] when there are no self-edges.

If there are some self-edges then expand out [\ref{eq:C2}] to
\begin{equation}
    \sum_{C_{2+}} \prod_{\gamma' \in C_{2+}} (-z^{|\gamma'|}w_{\gamma'}),
    \label{eq:C2+}
\end{equation}
where the sum is over collections $C_{2+}$ of disjoint elementary cycles formed by adding any 1-cycles to $C_2$, including the case of adding no 1-cycles.
Lastly, the contribution of the identity permutation is
\begin{equation}
    \prod_m (1-z w_{mm}) = 1 + \sum_{C_1} \prod_{\gamma \in C_1} (-z w_\gamma),
    \label{eq:C1}
\end{equation}
where the sum is over non-empty collections of disjoint 1-cycles.
Adding together [\ref{eq:C2+}] and [\ref{eq:C1}], we obtain the result [\ref{eq:cycleexp}] for the general case.

\end{document}